\documentclass[%
 reprint,
superscriptaddress,
 amsmath,amssymb,
 aps,
pra,
]{revtex4-2}

\usepackage{graphicx}
\usepackage{dcolumn}
\usepackage{bm}
\usepackage{mathtools}
\usepackage{comment} 
\usepackage{hyperref}
\usepackage{siunitx}
\usepackage[dvipsnames]{xcolor}

\usepackage{braket}
\hypersetup{
    colorlinks=true,
    citecolor=Blue,
    linkcolor=Blue,
    urlcolor=Blue
}

\usepackage{enumitem}

\usepackage{booktabs} 
\usepackage{array}    
\usepackage{amsmath}  
\usepackage{nicematrix}

\usepackage{xcolor}

\newcolumntype{C}[1]{>{\centering\arraybackslash}p{#1}}
\newcolumntype{L}[1]{>{\raggedright\arraybackslash}p{#1}}
\newcolumntype{R}[1]{>{\raggedleft\arraybackslash}p{#1}}

\newcommand{\onep}{\ensuremath{1^{T}}}
\newcommand{\ones}{\ensuremath{1^{*}}}
\newcommand{\oneps}{\ensuremath{1^{\dagger}}}

\begin{document}

\title{Compound symmetries and double antisymmetry groups \protect\\ in linear time-invariant photonic systems}

\author{Yunrui Wang}
\affiliation{Department of Electrical and Computer Engineering and Microelectronics Research Center, The University of Texas at Austin, Austin, Texas 78712, USA}

\author{Shiyu Li}
\affiliation{Department of Electrical and Computer Engineering and Microelectronics Research Center, The University of Texas at Austin, Austin, Texas 78712, USA}

\author{Cheng Guo}
\email[Contact author: ]{chengguo@utexas.edu}
\affiliation{Department of Electrical and Computer Engineering and Microelectronics Research Center, The University of Texas at Austin, Austin, Texas 78712, USA}

\received{17 April 2026; accepted 13 July 2026; published 3 August 2026}

\begin{abstract}
Symmetry is fundamental to photonic systems. External (spatial) symmetries and internal symmetries---Lorentz reciprocity, energy conservation, and time-reversal symmetry---constrain the electromagnetic response. Photonic systems can also possess compound symmetries that combine external and internal transformations, exemplified by parity-time ($\mathcal{PT}$) symmetry. However, a unified framework for general compound symmetries involving reciprocity, energy conservation, and time reversal remains lacking, leaving their classification and physical implications unexplored. In this paper, we present such a framework for linear photonic systems. We define compound transformations and symmetries, and derive their constraints on electromagnetic fields and scattering matrices. We show that internal, external, and compound symmetries are naturally described by the theory of double antisymmetry groups. This theory classifies linear time-invariant photonic systems into twelve symmetry categories, each imposing characteristic constraints on the electromagnetic response. We illustrate two representative categories with numerical examples of photonic crystal slabs and apply the theory to examine Kirchhoff's law of thermal radiation for a gyrotropic sphere. Our work provides a systematic foundation for analyzing and engineering symmetry in photonic systems.
\end{abstract}

\maketitle

\section{Introduction}\label{sec:introduction}

Symmetry is a central principle in physics.
A system is symmetric when it remains invariant under a transformation, and such invariance constrains its physical behavior~\cite{schwichtenberg2017}.
In photonics, two classes of symmetry play prominent roles.
External symmetries---associated with spatial transformations such as rotations, reflections, and translations---govern polarization selection rules, mode degeneracies, and band structures~\cite{okamoto2006,fan1998channel,rodriguez2018symmetry,meade2008photonic,cai2020symmetry}.
Internal symmetries, intrinsic to Maxwell's equations, impose three global constraints on linear electromagnetic systems: Lorentz reciprocity, energy conservation, and time-reversal symmetry~\cite{jinaukong1972,jackson1998classical,potton2004reciprocity,landau2013electrodynamics,guo2022reciprocity,guo2023a}.
In a recent work~\cite{guo2022internal}, we showed that these three constraints can be studied in a unified way in terms of fundamental matrix operations. We defined an internal transformation for each constraint, corresponding to the transpose, conjugate transpose, and conjugate of the system operator, respectively. These three internal transformations, together with the identity, form the Klein four-group $V_4$, and the internal-symmetry group of any linear photonic system is one of the five subgroups of $V_4$~\cite{guo2022internal}.

Beyond external and internal symmetries, photonic systems can also exhibit compound symmetries that combine external and internal transformations. A prominent example is parity-time ($\mathcal{PT}$) symmetry, the invariance under the composition of spatial inversion and time reversal~\cite{bender1998real}. In the unbroken $\mathcal{PT}$-symmetric phase, non-Hermitian photonic systems can still exhibit entirely real eigenvalue spectra~\cite{guo2009observation,longhi2009bloch,ruter2010observation}. Near $\mathcal{PT}$-symmetry-breaking points, such systems display unusual wave phenomena~\cite{ge2014parity,el2018non,zhao2018parity,cheng2022truncation,bai2026recent} that enable applications in lasing~\cite{chang2014parity,feng2014single,hodaei2014parity,fang2025}, coherent perfect absorption~\cite{chong2010coherent,wan2011time,longhi2011coherent}, optical sensing~\cite{wiersig2014enhancing,chen2017exceptional,anderson2023clarification,bai2023nonlinear,bai2024observation}, and unidirectional transmission~\cite{regensburger2012parity,lin2011unidirectional,peng2014parity}. Photonic systems may also possess other compound symmetries involving reciprocity or energy conservation, which constrain absorption, emission, and scattering in non-Hermitian and nonreciprocal systems~\cite{guo2022adjoint,ozaktas2026reciprocity,jin2022unitary,xu2023pseudo,li2024braiding}.

Despite these advances on specific compound symmetries, a general theory of compound symmetries involving reciprocity, energy conservation, and time reversal remains lacking. Three fundamental questions remain unanswered: what compound symmetries are possible in linear photonic systems, how do they classify photonic systems, and what constraints do they impose on electromagnetic fields and scattering matrices?

In this paper, we develop a unified theory of compound transformations and symmetries associated with reciprocity, energy conservation, and time reversal in linear photonic systems. We define compound transformations as compositions of these internal transformations with external transformations, and derive the constraints they impose on electromagnetic fields and scattering matrices. We show that internal, external, and compound symmetries are naturally described by double antisymmetry groups~\cite{heesch1930xix,cochran1952symmetry,shubnikov1951symmetry,belov1956color,zamorzaev1957simmetriya,belov19571651,zamorzaev1976theory,zamorzaev1978color,zamorzaev1980antisymmetry,zamorzaev1988generalized}, a mathematical framework that adjoins the anti-identity operations of the internal transformations to the spatial group of external transformations. This theory classifies linear time-invariant photonic systems into twelve symmetry categories, each imposing characteristic constraints on the electromagnetic response. We illustrate this classification with two photonic crystal slabs and apply the theory to examine Kirchhoff's law of thermal radiation for a gyrotropic sphere. Our results provide a systematic foundation for understanding and engineering symmetry in linear photonic systems.

The goal of this paper is to develop a comprehensive double-antisymmetry group theory for linear time-invariant photonic systems in a rigorous, systematic, and self-contained manner. As such, our framework incorporates many known results: (1) internal symmetries, as introduced in Ref.~\cite{guo2022internal}; (2) external symmetries under spatial transformations~\cite{serdyukov2001,tinder2008}; (3) reciprocity-based compound symmetries~\cite{guo2022adjoint,ozaktas2026reciprocity}; and (4) pseudo-Hermiticity~\cite{jin2022unitary,xu2023pseudo,li2024braiding}. Beyond providing a unified framework for these known results, our theory also predicts new results. First, we develop a comprehensive theory of energy-conservation-based compound symmetries. Second, we identify a broad family of systems that possess all three types of compound symmetries but no individual internal symmetry. Third, we provide a complete classification of all linear time-invariant systems with respect to these three types of compound symmetries. As an application of our group theory, we address a recent question concerning whether a gyrotropic sphere obeys or violates Kirchhoff’s law of thermal radiation. 

We also discuss several related works. References~\cite{hehl2005linear,tretyakov1998magnetoelectric} studied the decomposition and classification of local constitutive tensors for homogeneous bianisotropic media. Single-antisymmetry groups, also known as magnetic groups, have been used in such classifications~\cite{dmitriev1998constitutive,dmitriev1998degeneracy,yin1999reflection}. However, most existing works consider only the time-reversal operator~\cite{prudencio2015asymmetric,dmitriev1998constitutive,dmitriev1998degeneracy}. In contrast, our work uses double-antisymmetry groups to study the global properties of inhomogeneous systems containing bianisotropic media. Recent work~\cite{ozaktas2026reciprocity} established reciprocity-based compound symmetries for such inhomogeneous systems. In the present work, we further extend this framework to include energy-conservation-based compound symmetries. Exploring how our framework informs the tensor classifications developed in Refs.~\cite{hehl2005linear,tretyakov1998magnetoelectric} is an interesting direction for future research. Finally, our work is related to recent studies on the symmetry classification of non-Hermitian systems~\cite{kawabata2019symmetry}, but differs in scope and focus. Those studies concern the symmetries of effective Hamiltonians and their topological consequences, whereas our work focuses on the symmetries of photonic systems and their consequences for wave scattering. 

The rest of this paper is organized as follows: In Sec.~\ref{sec:review_internal_symmetry}, we review internal transformations and symmetries. In Sec.~\ref{sec:external_transformations}, we discuss external transformations and symmetries. In Sec.~\ref{sec:compound_transformations}, we define compound transformations and symmetries, and derive the fundamental relations of electromagnetic fields under these transformations. In Sec.~\ref{sec:S_matrix}, we derive the constraints on scattering matrices under external and compound transformations. In Sec.~\ref{sec:antisymmetry_group_theory}, we develop a unified group theory in terms of double antisymmetry groups, classify linear photonic systems into twelve categories, illustrate two categories and their symmetry constraints with numerical examples, and apply the theory to examine Kirchhoff's law for a gyrotropic sphere. We conclude in Sec.~\ref{sec:conclusion}.

\section{Internal transformations and symmetries}\label{sec:review_internal_symmetry}

We briefly review the fundamental internal symmetries of linear systems~\cite{guo2022internal}. A general ``original'' linear system is represented by the equation
\begin{equation}
   \text{original system:} \qquad \qquad \quad A x = b,
    \label{eq:orig-Ax=b}
\end{equation}
where $A$ is a complex matrix, and $x$ and $b$ are complex column vectors. We construct three related systems via elementary matrix operations:
\begin{align}
    &\text{transposed system:} &A^{T} x_{1} = b_{1},\label{eq:transposed-A}\\
    &\text{$c$-transposed system:} &A^{\dagger} x_{2} = b_{2}, \label{eq:ctransposed-A}\\
    &\text{conjugated system:} &A^{*} x_{3} = b_{3}. \label{eq:conjugated-A}
\end{align}
Here, “$c$-transpose” denotes the conjugate transpose, also known as the Hermitian conjugate or Hermitian adjoint, and “conjugate” denotes the complex conjugation~\cite{guo2022internal}. These three systems are related to the original by three fundamental identities:
\begin{align}
    x_{1}^{T}b - x^{T}b_{1} &= x_1^TAx-x^TA^Tx_1 = 0,\label{eq:id-transpose} \\
    x_{2}^{\dagger}b - \bigl(x^{\dagger}b_{2}\bigr)^{*} &= x_2^\dagger Ax-(x^\dagger A^\dagger x_2)^{*} = 0,\label{eq:id-c-transpose} \\
    A^{*}x^{*} &= b^{*},\label{eq:id-conjugate}
\end{align}
which reduce to symmetry constraints when $A$ is self-transpose ($A=A^T$), self-$c$-transpose ($A=A^\dagger$), or self-conjugate ($A=A^\ast$), respectively.

In this paper, we apply the general theory to linear time-invariant photonic systems. Here, an original system made of a local inhomogeneous dispersive bianisotropic medium is described by a $6\times6$ constitutive matrix field $C(\omega,\bm{r})$:
\begin{equation}\label{eq:C_matrix}
\begin{aligned}[b]
\begin{pmatrix}
\bm D(\omega,\bm r)\\[2pt] \bm B(\omega,\bm r)
\end{pmatrix}
&= C(\omega,\bm r)
\begin{pmatrix}
\bm E(\omega,\bm r)\\[2pt] \bm H(\omega,\bm r)
\end{pmatrix} \\
&=
\begin{pmatrix}
\varepsilon(\omega,\bm r) & \zeta(\omega,\bm r)\\
\eta(\omega,\bm r) & \mu(\omega,\bm r)
\end{pmatrix}
\begin{pmatrix}
\bm E(\omega,\bm r)\\[2pt] \bm H(\omega,\bm r)
\end{pmatrix}.
\end{aligned}
\end{equation}
Maxwell's equations for this system take a compact linear form analogous to Eq.~\eqref{eq:orig-Ax=b}:
\begin{equation}\label{eq:Maxwell_original_short}
M(\omega,\bm r)\,\Phi(\omega,\bm r)=a(\omega,\bm r),
\end{equation}
where the field and source vectors are
\begin{equation}
\Phi(\omega,\bm r) = 
\begin{pmatrix}
\bm E(\omega,\bm r)\\[2pt] i\bm H(\omega,\bm r)
\end{pmatrix},\quad
a(\omega,\bm r) = 
\begin{pmatrix}
i\bm J(\omega,\bm r)\\[2pt] \bm 0
\end{pmatrix},
\end{equation}
and the system operator is
\begin{equation}\label{eq:Maxwell_original_short_where-1}
M(\omega,\bm r) =  N_C(\omega,\bm r)+M_{\nabla}.
\end{equation}
The matrix $N_C(\omega,\bm r)$ encodes the material properties,
\begin{equation}\label{eq:def_N_C}
N_C(\omega,\bm r) = 
\begin{pmatrix}
-\omega\,\varepsilon(\omega,\bm r) & i\omega\,\zeta(\omega,\bm r)\\
-\,i\omega\,\eta(\omega,\bm r) & -\omega\,\mu(\omega,\bm r)
\end{pmatrix},
\end{equation}
while $M_{\nabla}$ is a system-independent differential operator,
\begin{equation}\label{eq:def_M_Nabla}
M_{\nabla} = 
\begin{pmatrix}
0 & \bm{\nabla}\times\\
\bm{\nabla}\times & 0
\end{pmatrix}.
\end{equation}

Applying the three transformations to this system yields three systems with constitutive matrix fields
\begin{align}
    &C_{1}(\omega,\bm{r}) = \begin{pmatrix}\varepsilon^{T}(\omega,\bm{r})&-\,\eta^{T}(\omega,\bm{r})\\ -\,\zeta^{T}(\omega,\bm{r})&\mu^{T}(\omega,\bm{r})\end{pmatrix},\label{eq:transposed-A-EM}\\
    &C_{2}(\omega,\bm{r}) = \begin{pmatrix}\varepsilon^{\dagger}(\omega,\bm{r})&\eta^{\dagger}(\omega,\bm{r})\\ \zeta^{\dagger}(\omega,\bm{r})&\mu^{\dagger}(\omega,\bm{r})\end{pmatrix}, \label{eq:ctransposed-A-EM}\\
    &C_{3}(\omega,\bm{r}) = \begin{pmatrix}\varepsilon^{*}(\omega,\bm{r})&-\,\zeta^{*}(\omega,\bm{r})\\ -\,\eta^{*}(\omega,\bm{r})&\mu^{*}(\omega,\bm{r})\end{pmatrix}.\label{eq:conjugated-A-EM}
\end{align}
Physically, the transposed system $C_1(\omega,\bm{r})$ is obtained from the original system $C(\omega,\bm{r})$ by reversing the bias magnetic field everywhere, the $c$-transposed system $C_2 (\omega,\bm{r})$ by interchanging loss and gain everywhere, and the conjugated system $C_3 (\omega,\bm{r})$ by applying both operations~\cite{guo2022internal}. The relations between these transformed systems and the original system yield three physical identities~\cite{guo2022internal}: the generalized reciprocity theorem~\cite{jinaukong1972,kong1970}, the modified mutual energy theorem~\cite{zhao2015}, and the generalized time-reversal theorem~\cite{guo2022internal}. These identities reduce to symmetry constraints when $C$ is invariant under each transformation: Lorentz reciprocity ($C=C_1$), energy conservation ($C = C_2$), and time-reversal symmetry ($C = C_3$). These internal symmetries impose constraints on the scattering matrix $S$, summarized in Table~\ref{tab:symmetries}.

\begin{table}[htbp]
    \caption{\label{tab:symmetries} Internal symmetries and $S$-matrix constraints~\cite{guo2022internal}.}
    \begin{ruledtabular}
        \begin{tabular}{lll}
            Internal symmetry & $C$-matrix  & $S$-matrix\\
            \hline\vspace{-8pt}\\
            Lorentz reciprocity & $C = C_1$ &  $S = S^{T}$ \\
            Energy conservation & $C = C_2$ & $S = (S^{\dagger})^{-1}$ \\
            Time-reversal symmetry & $C = C_3$ & $S = (S^{*})^{-1}$ \\
        \end{tabular}
    \end{ruledtabular}
\end{table}

The three internal transformations, denoted by $1^{T}$, $1^{\dagger}$, and $1^*$, combine with the identity operation $1$ to form the Klein four-group $V_{4} = \mathbb{Z}_2 \times \mathbb{Z}_2 = \{1, 1^{T}, 1^{\dagger}, 1^*\}$, where $\mathbb{Z}_2$ is the cyclic group of order 2 and $\times$ denotes the direct product. The internal-symmetry group of any linear photonic system is one of the five subgroups of $V_{4}$~\cite{guo2022internal}.

\section{External transformations and symmetries}\label{sec:external_transformations}

External transformations are geometric transformations applied to a linear photonic system. An external symmetry is an external transformation that leaves the system invariant. For a finite system, the external transformations are orthogonal transformations $\mathcal{O} \in \mathrm{O}(3)$, and all external symmetries form a point group, a subgroup of $\mathrm{O}(3)$. For a periodic system, the external transformations further include translations, glide planes, and screw axes, and all external symmetries form a space group, a subgroup of the Euclidean group $\mathrm{E}(3)$. Our discussion mainly addresses three-dimensional finite photonic systems; the analysis extends readily to infinite (periodic or aperiodic) and lower-dimensional systems.

\subsection{Mathematical background}\label{subsec:mathematical_backgrounds}

We begin with a brief review of orthogonal transformations. Each orthogonal transformation $\mathcal{O} \in \mathrm{O}(3)$ is represented by a $3\times3$ real orthogonal matrix $O$ satisfying $O^{-1} = O^T$ and $\det(O) = \pm 1$. The transformation $\mathcal{O}$ is a proper rotation when $\det(O) = 1$ and an improper rotation otherwise. The proper rotations form the subgroup $\mathrm{SO}(3) \subseteq \mathrm{O}(3)$.

The group $\mathrm{O}(3)$ acts on complex fields over $\mathbb{R}^3$, as summarized below, with unbarred and barred quantities denoting polar (true) and axial (pseudo) fields, respectively~\cite{jackson1998classical}.

\begin{enumerate}[wide, labelwidth=!, labelindent=0pt, label=(\arabic*)]
    \item Scalar fields:
    \begin{equation}
    c(\bm{r})\rightarrow c(O^{-1}\bm{r}), \quad \bar{c}(\bm{r})\rightarrow \det (O)\, \bar{c}(O^{-1}\bm{r}).
    \end{equation} 
    \item Vector fields:
    \begin{equation}\label{eq:transform_vector}
    v(\bm{r})\to Ov(O^{-1}\bm{r}), \quad \bar{v}(\bm{r})\to \det(O)\,O\bar{v}(O^{-1}\bm{r}).
   \end{equation} 
   \item Second-order tensor fields:
   \begin{equation}
    t(\bm{r})\to O t(O^{-1}\bm{r}) O^{-1}, \; \bar{t}(\bm{r})\to \det(O)\,O \bar{t}(O^{-1}\bm{r}) O^{-1}.
   \end{equation} 
\end{enumerate}

The curl operator $\bm{\nabla}\times$ acts on the transformed vector fields in Eq.~\eqref{eq:transform_vector} as
\begin{align}
\label{eq:curl_theorem_v}
\bm{\nabla} \times [Ov(O^{-1}\bm{r})] = \det(O)\,O\,(\bm{\nabla} \times v)(O^{-1}\bm{r}), \\
\label{eq:curl_theorem_vbar}
\bm{\nabla} \times [\det(O)\,O\bar{v}(O^{-1}\bm{r})] = O\,(\bm{\nabla} \times \bar{v})(O^{-1}\bm{r}),
\end{align}
with the proof given in Appendix~\ref{appendix:proof_curl}. Since the curl of a polar vector field is axial and vice versa, Eqs.~\eqref{eq:curl_theorem_v} and~\eqref{eq:curl_theorem_vbar} show that the curl operator commutes with orthogonal transformations.

\subsection{Orthogonal transformations of linear photonic systems}

We derive how an orthogonal transformation $\mathcal{O}\in \mathrm{O}(3)$ transforms a linear photonic system described by the constitutive matrix field $C(\omega,\bm{r})$, treating proper and improper rotations in turn:

\begin{enumerate}[wide, labelwidth=!, labelindent=0pt, label=(\arabic*)]
    \item \emph{Proper rotations:} Consider a proper rotation $\mathcal{O} = \mathcal{R}_{\bm{\theta}}\in \mathrm{SO}(3)$, where $\bm{\theta}$ specifies the rotation axis and angle. We denote the corresponding $3\times3$ orthogonal matrix as $O = R_{\bm{\theta}}$ with $\det (R_{\bm{\theta}}) = 1$. Under $\mathcal{R}_{\bm{\theta}}$, the transformed system is described by the constitutive matrix~\cite{serdyukov2001,tinder2008}
    \begin{equation}\label{eq:C_R}
    \hspace{18pt}
        \begin{aligned}[b]
            &\,C_{R_{\bm{\theta}}}(\omega,\bm{r}) \\ &=\! \begin{pmatrix}
            R_{\bm{\theta}}\,\varepsilon(\omega, {R}^{-1}_{\bm{\theta}}\bm{r})\,{R}^{-1}_{\bm{\theta}} & \!R_{\bm{\theta}}\,\zeta(\omega, {R}^{-1}_{\bm{\theta}}\bm{r})\,{R}^{-1}_{\bm{\theta}} \\ R_{\bm{\theta}}\,\eta(\omega, {R}^{-1}_{\bm{\theta}}\bm{r})\,{R}^{-1}_{\bm{\theta}} & \!R_{\bm{\theta}}\,\mu(\omega, {R}^{-1}_{\bm{\theta}}\bm{r})\,{R}^{-1}_{\bm{\theta}}
            \end{pmatrix}.\!
        \end{aligned}
    \end{equation}
$\mathcal{R}_{\bm{\theta}}$ is a symmetry of the system if $C(\omega,\bm{r}) = C_{R_{\bm{\theta}}}(\omega,\bm{r})$.
    
    \item \emph{Improper rotations:} Consider an improper rotation $\mathcal{O} = \mathcal{Q}_{\bm{\theta}} = \mathcal{I}\mathcal{R}_{\bm{\theta}}\in \mathrm{O}(3)\backslash \mathrm{SO}(3)$, where $\mathcal{I}$ is spatial inversion and $\mathcal{R}_{\bm{\theta}}$ is a proper rotation. We denote the corresponding $3\times3$ orthogonal matrix as $O = Q_{\bm{\theta}}$ with $\det (Q_{\bm{\theta}}) = -1$. Under $\mathcal{Q}_{\bm{\theta}}$, the transformed system is described by the constitutive matrix~\cite{serdyukov2001,tinder2008}
    \begin{equation}\label{eq:C_Q}
        \hspace{18pt}
        \begin{aligned}[b]
            &C_{Q_{\bm{\theta}}}(\omega,\bm{r})\\ &= \begin{pmatrix}
            Q_{\bm{\theta}}\,\varepsilon(\omega, {Q}^{-1}_{\bm{\theta}}\bm{r})\,{Q}^{-1}_{\bm{\theta}} &\! -Q_{\bm{\theta}}\,\zeta(\omega, {Q}^{-1}_{\bm{\theta}}\bm{r})\,{Q}^{-1}_{\bm{\theta}} \\ -Q_{\bm{\theta}}\,\eta(\omega, {Q}^{-1}_{\bm{\theta}}\bm{r})\,{Q}^{-1}_{\bm{\theta}} & \!Q_{\bm{\theta}}\,\mu(\omega, {Q}^{-1}_{\bm{\theta}}\bm{r})\,{Q}^{-1}_{\bm{\theta}}
            \end{pmatrix}, 
        \end{aligned}
    \end{equation}
    where the minus signs in the off-diagonal blocks arise because $\zeta$ and $\eta$ are axial tensors. $\mathcal{Q}_{\bm{\theta}}$ is a symmetry of the system if $C(\omega,\bm{r}) = C_{Q_{\bm{\theta}}}(\omega,\bm{r})$.
\end{enumerate}

These two cases admit a unified treatment. Consider $\mathcal{O}\in \mathrm{O}(3)$ represented by a $3\times3$ orthogonal matrix $O$, and define a $6\times6$ real orthogonal matrix
\begin{equation}
   W =
   \begin{pmatrix}
   O & 0 \\ 0 & \det (O)\, O
   \end{pmatrix}.
\end{equation}
Under $\mathcal{O}$, the transformed system is described by
\begin{align}\label{eq:O_unified}
&C_{O} (\omega, \bm{r})  = W  C(\omega, O^{-1}\bm{r})W^{-1} \notag \\ &=\begin{pmatrix}
\hspace{-2em} O\varepsilon(\omega, O^{-1}\bm{r})O^{-1} & \hspace{-2.0em}\det(O)  O\zeta(\omega, O^{-1}\bm{r})O^{-1} \\ \det(O) O\eta(\omega, O^{-1}\bm{r}){O}^{-1} & O\mu(\omega, {O}^{-1}\bm{r})O^{-1}
\end{pmatrix}.
\end{align} 
Equation~\eqref{eq:O_unified} reduces to Eq.~\eqref{eq:C_R} when $\det (O) = 1$ and to Eq.~\eqref{eq:C_Q} when $\det (O) = -1$. Hereafter, the subscript $O$ denotes a transformed quantity obtained from an original quantity under $\mathcal{O}$.

The transformed system satisfies Maxwell's equations
\begin{equation}\label{eq:Maxwell_O_short}
 M_{O}(\omega,\bm{r})\, \Phi'(\omega,\bm{r}) = a'(\omega,\bm{r}),   
\end{equation}
where the transformed field and source vectors are
\begin{equation}
\Phi'(\omega,\bm{r}) =
   \begin{pmatrix}
   \bm{E}'(\omega, \bm{r}) \\ i\bm{H}'(\omega, \bm{r})
   \end{pmatrix}, \,
    a'(\omega,\bm{r}) =
    \begin{pmatrix}
    i\bm{J}'(\omega, \bm{r}) \\ \bm{0}
    \end{pmatrix},
\end{equation}
and the transformed system operator is
\begin{equation}
    M_O(\omega,\bm{r}) = N_{C_O}(\omega,\bm{r}) + M_\nabla,
\end{equation}
with
\hspace{-2pt}
\begin{align}
   &N_{C_O}(\omega,\bm{r}) = W N_C(\omega,O^{-1}\bm{r}) W^{-1} \notag\\
   &=
   \begin{pmatrix}
   \hspace{-3.5em} -\omega O\varepsilon(\omega, O^{-1}\bm{r})O^{-1}\! & \hspace{-2.5em} i\omega \det(O) O\zeta(\omega, O^{-1}\bm{r})O^{-1} \\ -i\omega\det(O) O\eta(\omega, O^{-1}\bm{r})O^{-1}\! & \hspace{-3pt}\! -\omega O\mu(\omega, O^{-1}\bm{r})O^{-1}
   \end{pmatrix}.
\end{align}

We define the transformation $C(\omega,\bm{r}) \to C_{O}(\omega,\bm{r})$ as an external transformation under $\mathcal{O}$. $\mathcal{O}$ is an external symmetry of the system if $C(\omega,\bm{r}) = C_{O}(\omega,\bm{r})$. In that case, $N_C(\omega,\bm{r}) = N_{C_O}(\omega,\bm{r})$ implies $M(\omega,\bm{r}) = M_O(\omega,\bm{r})$, so Maxwell's equations are invariant under $\mathcal{O}$.

\subsection{Fundamental relations under orthogonal transformations}

We derive the relation between the original and transformed systems under an external transformation.

Consider an orthogonal transformation $\mathcal{O} \in \mathrm{O}(3)$, which transforms an original system $C(\omega,\bm{r})$ into a transformed system $C_O(\omega,\bm{r})$. By Eq.~\eqref{eq:transform_vector}, $\mathcal{O}$ transforms the original fields $(\bm{E},\bm{H},\bm{J})$ into the transformed fields $(\bm{E}_O,\bm{H}_O,\bm{J}_O)$:
\begin{align}\label{eq:fields_O_E}
    \bm{E}_O(\omega,\bm{r}) &= O\bm{E}(\omega, O^{-1}\bm{r}),\\  \bm{H}_O(\omega,\bm{r}) &= \det(O) O\bm{H}(\omega, O^{-1}\bm{r}),\label{eq:fields_O_H} \\ \bm{J}_O(\omega,\bm{r}) &= O\bm{J}(\omega, O^{-1}\bm{r}), \label{eq:fields_O_J} 
\end{align}
since $\bm{E}$ and $\bm{J}$ are polar vector fields while $\bm{H}$ is an axial-vector field. More compactly, 
\begin{equation}
    \begin{aligned}[b]
&\Phi_O(\omega,\bm{r}) =
   \begin{pmatrix}
   \bm{E}_O(\omega, \bm{r}) \\ i\bm{H}_O(\omega, \bm{r})
   \end{pmatrix} \\ &= 
   \begin{pmatrix}
   O\bm{E}(\omega, O^{-1}\bm{r}) \\ i\det(O)\, O\bm{H}(\omega, O^{-1}\bm{r})
   \end{pmatrix}
   = W \Phi(\omega,O^{-1}\bm{r}),
   \end{aligned}
\end{equation}
\begin{equation}
\begin{aligned}[b]
    a_O(\omega,\bm{r}) &=
    \begin{pmatrix}
    i\bm{J}_O(\omega, \bm{r}) \\ \bm{0}
    \end{pmatrix} \\ &= 
    \begin{pmatrix}
    i\,O\bm{J}(\omega, O^{-1}\bm{r}) \\ \bm{0}
    \end{pmatrix}
    = W a(\omega,O^{-1}\bm{r}).
\end{aligned}
\end{equation}

We now state the fundamental relation imposed by an orthogonal transformation: the fields $(\bm{E},\bm{H},\bm{J})$ satisfy Maxwell's equations for the original system [Eq.~\eqref{eq:Maxwell_original_short}] if and only if the transformed fields $(\bm{E}_O,\bm{H}_O,\bm{J}_O)$ satisfy Maxwell's equations for the transformed system [Eq.~\eqref{eq:Maxwell_O_short}]. Mathematically, 
\begin{equation}\label{eq:external_symmetry_short}
\begin{aligned}[b]
    &M(\omega,\bm{r}) \, \Phi(\omega,\bm{r})  = a(\omega,\bm{r}) \\& \!\iff \! M_{O}(\omega,\bm{r})\, \Phi_{O}(\omega,\bm{r}) = a_{O}(\omega,\bm{r}). 
\end{aligned}
\end{equation}

Equation~\eqref{eq:external_symmetry_short} relates a set of sources and fields in the original system to the transformed sources and fields in the transformed system. When $\mathcal{O}$ is an external symmetry, Eq.~\eqref{eq:external_symmetry_short} relates a pair of sources and responses in that single system.

We prove Eq.~\eqref{eq:external_symmetry_short} in two steps. We first show that the left-hand side implies the right-hand side:
\begin{equation}    
\label{eq:MoPhiO_M_nabla_V}
    \begin{aligned}[b]
    & M_{O}(\omega,\bm{r})\, \Phi_{O}(\omega,\bm{r}) \\
     &= [W N_C(\omega,O^{-1}\bm{r}) W^{-1} + M_\nabla] W \Phi(\omega,O^{-1}\bm{r}) \\
     &= W N_C(\omega,O^{-1}\bm{r})  \Phi(\omega,O^{-1}\bm{r}) + M_\nabla W \Phi(\omega,O^{-1}\bm{r}).
    \end{aligned}
\end{equation}
We have
\begin{align}
        &M_\nabla W \Phi(\omega,O^{-1}\bm{r}) \nonumber \\
        &=  \begin{pmatrix}
         0  & \bm{\nabla}\times \\ \bm{\nabla}\times & 0
         \end{pmatrix} 
         \begin{pmatrix}
           O & 0 \\ 0 & \det (O)\, O
           \end{pmatrix}
         \begin{pmatrix}
           \bm{E}(\omega, O^{-1}\bm{r}) \\ i\bm{H}(\omega, O^{-1}\bm{r})
           \end{pmatrix} \nonumber \\
         &= \begin{pmatrix}
           i\bm{\nabla} \times[\det(O)\, O\bm{H}(\omega, O^{-1}\bm{r})] 
           \\
           \bm{\nabla} \times [O\bm{E}(\omega, O^{-1}\bm{r})]  
           \end{pmatrix} \nonumber \\
        &= \begin{pmatrix}
           i\,O (\bm{\nabla} \times \bm{H}) (\omega, O^{-1}\bm{r})
           \\
           \det(O)\, O (\bm{\nabla} \times \bm{E})(\omega, O^{-1}\bm{r})  
           \end{pmatrix}   \label{eq:proof_O_nabla}\\
        &= \begin{pmatrix}
           O & 0 \\ 0 & \det (O)\, O
           \end{pmatrix} \begin{pmatrix}
         0  & \bm{\nabla}\times \\ \bm{\nabla}\times & 0
         \end{pmatrix}  
         \begin{pmatrix}
         \bm{E}(\omega, O^{-1}\bm{r})  \\ i\bm{H}(\omega, O^{-1}\bm{r}) 
         \end{pmatrix} \nonumber \\ 
         &= W M_\nabla  \Phi(\omega,O^{-1}\bm{r}),\label{eq:V_M_nabla} 
    \end{align}
where we used Eqs.~\eqref{eq:curl_theorem_v} and~\eqref{eq:curl_theorem_vbar} in Eq.~\eqref{eq:proof_O_nabla}. Substituting Eq.~\eqref{eq:V_M_nabla} into Eq.~\eqref{eq:MoPhiO_M_nabla_V} gives
\begin{align}
M_{O}(\omega,\bm{r})\, \Phi_{O}(\omega,\bm{r}) &=
W [N_C(\omega,O^{-1}\bm{r})+ M_\nabla]  \Phi(\omega,O^{-1}\bm{r}) \nonumber\\
&= W M(\omega,O^{-1}\bm{r}) \Phi(\omega,O^{-1}\bm{r}) \label{eq:proof_VMPhi}\\
&= W a(\omega,O^{-1}\bm{r})
= a_O(\omega,\bm{r}).\label{eq:proof_Va}
\end{align} 

To show the converse, we apply the inverse transformation $\mathcal{O}^{-1}$, represented by $O^{-1}= O^T$, which maps the transformed system back to the original system; the same procedure then completes the proof.

\section{Compound transformations and symmetries}\label{sec:compound_transformations}

Having discussed internal and external transformations and symmetries separately, we now turn to compound transformations and symmetries. A compound transformation is a composition of an internal and an external transformation. A familiar example is the parity-time transformation, which composes spatial inversion with time reversal. As we show, linear photonic systems admit many other compound transformations and symmetries, each with its own physical implications.

\subsection{Definition and the commutative property}\label{subsec:commutative}

We define compound transformations of linear photonic systems. We have defined three internal transformations $C(\omega,\bm{r}) \to C_k(\omega,\bm{r})$, $k=1,2,3$, and external transformations $C(\omega,\bm{r}) \to C_O(\omega,\bm{r})$, $O \in \mathrm{O}(3)$. A compound transformation, denoted $C(\omega,\bm{r}) \to C_{k,O}(\omega,\bm{r})$, is obtained by applying first an internal and then an external transformation: $C(\omega,\bm{r}) \to C_{k}(\omega,\bm{r}) \to C_{k,O}(\omega,\bm{r})$.

As we prove below, any internal transformation commutes with any external transformation. The order in the definition of a compound transformation therefore does not matter: applying first the external and then the internal transformation yields the same result, $C(\omega,\bm{r}) \to C_{O}(\omega,\bm{r}) \to C_{O,k}(\omega,\bm{r})=C_{k,O}(\omega,\bm{r})$.

As an illustration, we prove this commutativity for the transformation $C(\omega,\bm{r}) \to C_{1,O}(\omega,\bm{r})$. Applying first the internal and then the external transformation:
    \begin{align}
&C(\omega,\bm{r}) =     \begin{pmatrix}
\varepsilon(\omega,\bm{r}) &\zeta(\omega,\bm{r}) \\ \eta(\omega,\bm{r}) &\mu(\omega,\bm{r}) 
\end{pmatrix} \notag\\
&\quad \downarrow \notag \\
&C_1(\omega,\bm{r}) =     \begin{pmatrix}
\varepsilon^T(\omega,\bm{r}) &-\eta^T(\omega,\bm{r}) \\ -\zeta^T(\omega,\bm{r}) &\mu^T(\omega,\bm{r}) 
\end{pmatrix} \notag\\
&\quad \downarrow \notag \\
&C_{1,O}(\omega,\bm{r})= \nonumber\\
&\!\!\begin{pmatrix}\!
O\varepsilon^T(\omega, O^{-1}\bm{r})O^{-1} \!&\!\!\!\!\!\!\! -\det(O)  O\eta^T(\omega, O^{-1}\bm{r})O^{-1} \!\\\label{eq:C1O_result} \!-\det(O) O\zeta^T(\omega, O^{-1}\bm{r}){O}^{-1} \!&\! O\mu^T(\omega, {O}^{-1}\bm{r})\,O^{-1}\!
\end{pmatrix}
\end{align}
Applying first the external and then the internal transformation:
\begin{align}
&C(\omega,\bm{r}) =     \begin{pmatrix}
\varepsilon(\omega,\bm{r}) &\zeta(\omega,\bm{r}) \\ \eta(\omega,\bm{r}) &\mu(\omega,\bm{r}) 
\end{pmatrix} \notag\\
&\quad \downarrow \notag \\
&C_O(\omega,\bm{r}) = \nonumber\\   & \begin{pmatrix}
O\varepsilon(\omega, O^{-1}\bm{r})O^{-1} & \det(O)  O\zeta(\omega, O^{-1}\bm{r})O^{-1} \\ \det(O) O\eta(\omega, O^{-1}\bm{r}){O}^{-1} & O\mu(\omega, {O}^{-1}\bm{r})\,O^{-1}
\end{pmatrix} \notag\\
&\quad \downarrow \notag \\
&C_{O,1}(\omega,\bm{r}) = \nonumber\\
&\!\!\!\left(\begin{aligned}
[O\varepsilon(\omega, O^{-1}\bm{r})O^{-1}]^T\qquad &\!\!\!\!\! -[\det(O)  O\eta(\omega, O^{-1}\bm{r})O^{-1}]^T \\\! -[\det(O) O\zeta(\omega, O^{-1}\bm{r}){O}^{-1}]^T \!&\qquad [O\mu(\omega, {O}^{-1}\bm{r})\,O^{-1}]^T
\end{aligned}\right) \nonumber
\\ 
&= \!\!\left(\begin{aligned}
    O\varepsilon^T(\omega, O^{-1}\bm{r})O^{-1} \qquad&\!\!\!\! -\det(O)  O\eta^T(\omega, O^{-1}\bm{r})O^{-1} \\\! -\det(O) O\zeta^T(\omega, O^{-1}\bm{r}){O}^{-1} \!\!& \qquad O\mu^T(\omega, {O}^{-1}\bm{r})\,O^{-1}
\end{aligned}\right)\label{eq:C1O_final}
\end{align}
where we used $O^{-1} = O^T$ to obtain Eq.~\eqref{eq:C1O_final}. Comparing Eqs.~\eqref{eq:C1O_result} and~\eqref{eq:C1O_final} gives $C_{O,1}(\omega,\bm{r})=C_{1,O}(\omega,\bm{r})$.

Similarly, one can show $C_{O,2}(\omega,\bm{r})=C_{2,O}(\omega,\bm{r})$ and $C_{O,3}(\omega,\bm{r})=C_{3,O}(\omega,\bm{r})$. See Appendix~\ref{appendix:proof_commutative} for the complete proof, which utilizes the properties of orthogonal transformations: $O^{-1} = O^\dagger$ and $O = O^*$, respectively. We have thus proved that any internal transformation commutes with any external transformation.

\subsection{Three classes of compound transformations}\label{subsec:three_class_compound}

Compound transformations fall into three classes according to the internal transformation involved. Those involving the internal transformation of $X$ are called the compound transformations of $X$. We discuss each class in detail:
\begin{enumerate}[wide, labelwidth=!, labelindent=0pt, label=(\arabic*)]
    \item \emph{Compound transposed system:} Starting from the externally transformed system described by Eq.~\eqref{eq:Maxwell_O_short}, we take its transpose to obtain the compound transposed system
    \begin{equation}\label{eq:Maxwell_compound_transpose}
 M_{O}^T(\omega,\bm{r})\, \Phi'_1(\omega,\bm{r}) = a'_1(\omega,\bm{r}), 
    \end{equation}
where the field and source vectors are
\begin{equation}
\Phi_1'(\omega,\bm{r}) = 
    \begin{pmatrix}
    \bm{E}'_1(\omega,\bm{r}) \\ i\bm{H}'_1(\omega,\bm{r})
    \end{pmatrix}, \,
    a'_1(\omega,\bm{r}) =
    \begin{pmatrix}
    i\bm{J}'_1(\omega,\bm{r}) \\ \bm{0}
    \end{pmatrix},
\end{equation}
and the system operator is
\begin{equation}
M_O^T(\omega,\bm{r}) =    N_{C_{1,O}}(\omega,\bm{r}) + M_\nabla,  
\end{equation}
with
\begin{align}\label{eq:def_N_C_1O}
    & N_{C_{1,O}}(\omega,\bm{r}) = \nonumber\\
&\hspace{-1.5em}\begin{pmatrix}\hspace{-2.5em}
-\omega O\varepsilon^T(\omega, O^{-1}\bm{r})O^{-1} \!&\hspace{-2.5em} -i\omega \det(O) O\eta^T(\omega, O^{-1}\bm{r})O^{-1} \!\\ \!i\omega\det(O) O\zeta^T(\omega, O^{-1}\bm{r})O^{-1} \!&\! -\omega O\mu^T(\omega, O^{-1}\bm{r})O^{-1}\!
\end{pmatrix}.
\end{align}
The compound transposed system is thus described by the constitutive matrix 
\begin{align}\label{eq:C1O}
    &C_{1,O}(\omega,\bm{r})= \nonumber\\
&\hspace{-1.5em}\begin{pmatrix}\hspace{-3em} 
O\varepsilon^T(\omega, O^{-1}\bm{r})O^{-1} \!&\hspace{-3em} -\det(O)  O\eta^T(\omega, O^{-1}\bm{r})O^{-1} \!\\ \!-\det(O) O\zeta^T(\omega, O^{-1}\bm{r}){O}^{-1} \!&\! O\mu^T(\omega, {O}^{-1}\bm{r})O^{-1}
\end{pmatrix}
\end{align}
which is identical to Eqs.~\eqref{eq:C1O_result} and~\eqref{eq:C1O_final}. 

We define the class of transformations $C(\omega, \bm{r}) \to C_{1,O}(\omega, \bm{r})$, $O\in \mathrm{O}(3)$, as the compound transformations of reciprocity. A system satisfying $C(\omega,\bm{r}) = C_{1,O}(\omega,\bm{r})$ is called compound reciprocal.

\item \emph{Compound $c$-transposed system:} Starting from the externally transformed system described by Eq.~\eqref{eq:Maxwell_O_short}, we take its $c$-transpose to obtain the compound $c$-transposed system
    \begin{equation}\label{eq:Maxwell_compound_c-transpose}
 M_{O}^\dagger(\omega,\bm{r})\, \Phi'_2(\omega,\bm{r}) = a'_2(\omega,\bm{r}), 
    \end{equation}
where the field and source vectors are
\begin{equation}
\Phi_2'(\omega,\bm{r}) = 
    \begin{pmatrix}
    \bm{E}'_2(\omega,\bm{r}) \\ i\bm{H}'_2(\omega,\bm{r})
    \end{pmatrix}, \,
    a'_2(\omega,\bm{r}) =
    \begin{pmatrix}
    i\bm{J}'_2(\omega,\bm{r}) \\ \bm{0}
    \end{pmatrix},
\end{equation}
and the system operator is
\begin{equation}
M_O^\dagger(\omega,\bm{r}) =  N_{C_{2,O}}(\omega,\bm{r}) + M_\nabla,   
\end{equation}
with
\begin{align}\label{eq:def_N_C_2O}
    & N_{C_{2,O}}(\omega,\bm{r}) =  \nonumber\\
&\hspace{-1.5em}\begin{pmatrix}\hspace{-2.5em}
-\omega O\varepsilon^\dagger(\omega, O^{-1}\bm{r})O^{-1} \!&\hspace{-2.5em} i\omega \det(O) O\eta^\dagger(\omega, O^{-1}\bm{r})O^{-1} \!\\ \!-i\omega \det(O) O\zeta^\dagger(\omega, O^{-1}\bm{r})O^{-1} \!&\! -\omega O\mu^\dagger(\omega, O^{-1}\bm{r})O^{-1}\!
\end{pmatrix}.
\end{align}
The compound $c$-transposed system is thus described by the constitutive matrix 
\begin{align}\label{eq:C2O}
    &C_{2,O}(\omega,\bm{r})= \nonumber\\
&\hspace{-1em}\begin{pmatrix}\hspace{-2em} 
O\varepsilon^\dagger(\omega, O^{-1}\bm{r})O^{-1} \!&\hspace{-2em} \det(O)  O\eta^\dagger(\omega, O^{-1}\bm{r})O^{-1} \!\\ \det(O) O\zeta^\dagger(\omega, O^{-1}\bm{r}){O}^{-1} \!& O\mu^\dagger(\omega, {O}^{-1}\bm{r})O^{-1}
\end{pmatrix}.
\end{align}

We define the class of transformations $C(\omega, \bm{r}) \to C_{2,O}(\omega, \bm{r})$, $O\in \mathrm{O}(3)$, as the compound transformations of energy conservation. A system satisfying $C(\omega,\bm{r}) = C_{2,O}(\omega,\bm{r})$ is called compound energy-conserving.

\item \emph{Compound conjugated system:} Starting from the externally transformed system described by Eq.~\eqref{eq:Maxwell_O_short}, we take its conjugate to obtain the compound conjugated system
    \begin{equation}\label{eq:Maxwell_compound_conjugate}
 M_{O}^*(\omega,\bm{r})\, \Phi'_3(\omega,\bm{r}) = a'_3(\omega,\bm{r}),
    \end{equation}
where the field and source vectors are
\begin{equation}
\Phi_3'(\omega,\bm{r}) = 
    \begin{pmatrix}
    \bm{E}'_3(\omega,\bm{r}) \\ i\bm{H}'_3(\omega,\bm{r})
    \end{pmatrix},
    \,
    a'_3(\omega,\bm{r}) =
    \begin{pmatrix}
    i\bm{J}'_3(\omega,\bm{r}) \\ \bm{0}
    \end{pmatrix},
\end{equation}
and the system operator is
\begin{equation}
M_O^{*}(\omega,\bm{r}) =  N_{C_{3,O}}(\omega,\bm{r}) + M_\nabla,   
\end{equation}
with
\begin{align}\label{eq:def_N_C_3O}
    &N_{C_{3,O}}(\omega,\bm{r})= \nonumber\\
&\hspace{-1.5em}\begin{pmatrix}\hspace{-2.5em}
-\omega O\varepsilon^*(\omega, O^{-1}\bm{r})O^{-1} \!&\hspace{-2.5em} -i\omega \det(O) O\zeta^*(\omega, O^{-1}\bm{r})O^{-1} \!\\ \!i\omega \det(O) O\eta^*(\omega, O^{-1}\bm{r})O^{-1} \!&\! -\omega O\mu^*(\omega, O^{-1}\bm{r})O^{-1}\!
\end{pmatrix}.
\end{align}

The compound conjugated system is thus described by the constitutive matrix 
\begin{align}\label{eq:C3O}
    &C_{3,O}(\omega,\bm{r})= \nonumber\\
&\hspace{-1em}\begin{pmatrix}\hspace{-2em} 
O\varepsilon^*(\omega, O^{-1}\bm{r})O^{-1} \!&\hspace{-2em} -\det(O)  O\zeta^*(\omega, O^{-1}\bm{r})O^{-1}  \!\\ -\det(O) O\eta^*(\omega, O^{-1}\bm{r}){O}^{-1} \!& O\mu^*(\omega, {O}^{-1}\bm{r})O^{-1}
\end{pmatrix}.
\end{align}

We define the class of transformations $C(\omega, \bm{r}) \to C_{3,O}(\omega, \bm{r})$, $O\in \mathrm{O}(3)$, as the compound transformations of time reversal. A system satisfying $C(\omega,\bm{r}) = C_{3,O}(\omega,\bm{r})$ is called compound time-reversal symmetric. When $\mathcal{O}$ is spatial inversion, this symmetry is known as parity-time symmetry. 
\end{enumerate}

\begin{figure}[htbp]
    \centering
    \includegraphics[width=0.95\linewidth]{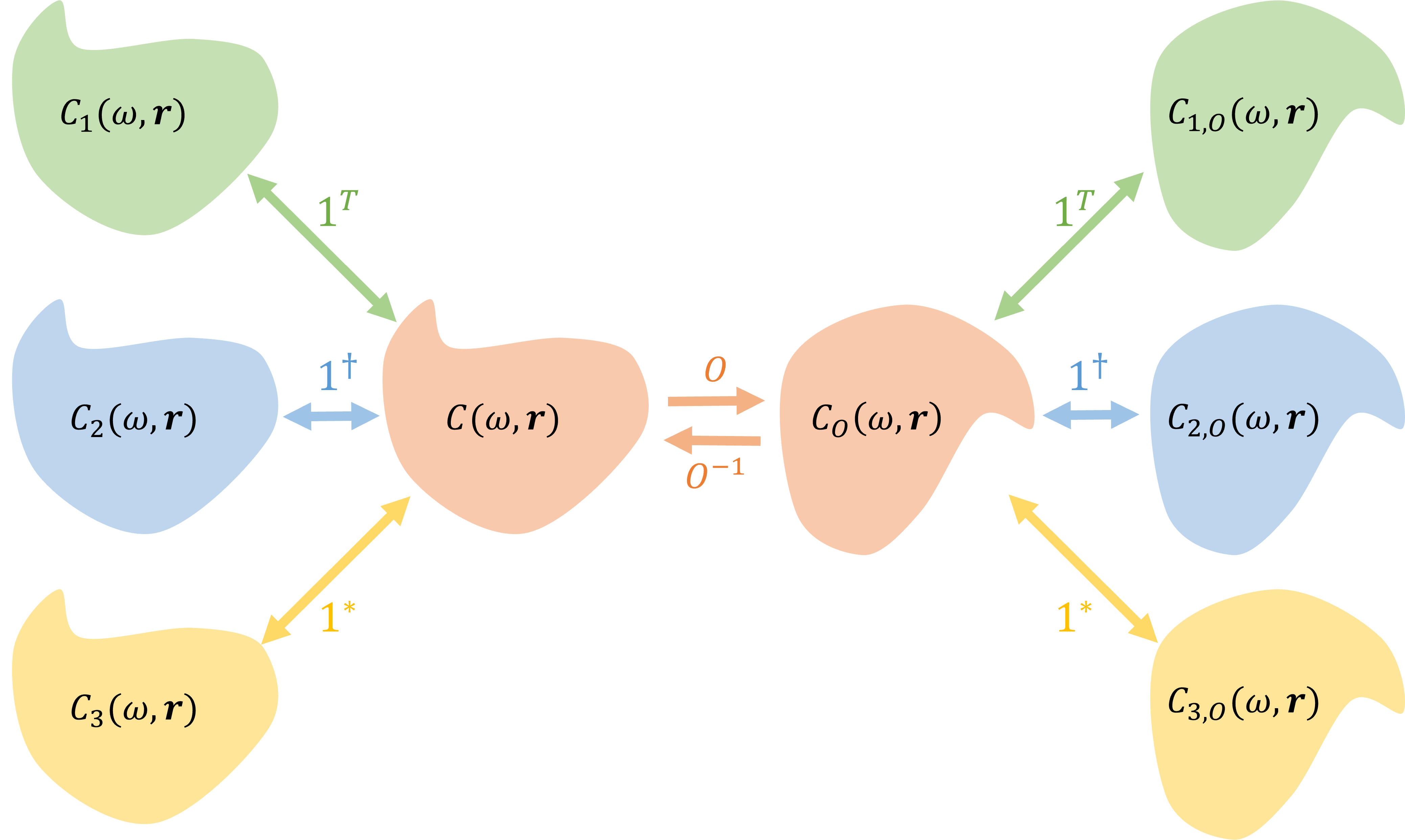}
     \caption{Scheme of transformed systems under internal, external, and compound transformations. The internal transformations $1^T$ (green), $1^\dagger$ (blue), and $1^*$ (yellow) map the original system $C(\omega,\bm{r})$ to the internally transformed systems $C_k(\omega,\bm{r})$, $k=1,2,3$. The external transformation $\mathcal{O}\in \mathrm{O}(3)$ maps $C(\omega,\bm{r})$ to the externally transformed system $C_O(\omega,\bm{r})$. Applying the internal transformations to $C_O(\omega,\bm{r})$ gives the compound transformed systems $C_{k,O}(\omega,\bm{r})$, $k=1,2,3$.}
\label{fig:scheme_systems}
\end{figure}

We have now defined transformed systems under internal, external, and compound transformations, summarized in Fig.~\ref{fig:scheme_systems}. Starting from an original system $C(\omega,\bm{r})$, the internal transformations $1^T$, $1^\dagger$, and $1^*$ give the internally transformed systems $C_k(\omega,\bm{r})$, $k=1,2,3$, respectively; the external transformation $\mathcal{O} \in \mathrm{O}(3)$ gives the externally transformed system $C_O(\omega,\bm{r})$. Applying $1^T$, $1^\dagger$, and $1^*$ to $C_O(\omega,\bm{r})$ gives the compound transformed systems $C_{k,O}(\omega,\bm{r})$, respectively.

\subsection{Relations under compound transformations}
\label{subsec:relation_compound}

We now derive the fundamental relations of electromagnetic fields between the original and transformed systems under compound transformations. We have derived the relations under internal and external transformations separately. Since a compound transformation composes an internal and an external transformation, we combine the previous results to obtain the desired relations.

We use the externally transformed system described by Eq.~\eqref{eq:Maxwell_O_short} as an auxiliary system. The fundamental relation between the original and auxiliary systems is given by Eq.~\eqref{eq:external_symmetry_short}, and the auxiliary fields $(\bm{E}_O,\bm{H}_O,\bm{J}_O)$ are determined from $(\bm{E},\bm{H},\bm{J})$ by Eqs.~\eqref{eq:fields_O_E}-\eqref{eq:fields_O_J}.
\begin{enumerate}[wide, labelwidth=!, labelindent=0pt, label=(\arabic*)]
\item \emph{Original and compound transposed systems:} The compound transposed system [Eq.~\eqref{eq:Maxwell_compound_transpose}] is the transpose of the auxiliary system, so they are related by the generalized reciprocity theorem [Eq.~(42) in Ref.~\cite{guo2022internal}]:
    \begin{equation}\label{eq:compound_reciprocity_integral}
    \begin{aligned}[b]
        &\oint_S (\bm{E}_O\times\bm{H}'_1\!-\!\bm{E}'_1\times\bm{H}_O) \cdot \bm{\mathrm{d}S} \\&= \int_V (\bm{E}'_1\cdot\bm{J}_O-\bm{E}_O\cdot\bm{J}'_1) \,\mathrm{d}V
    \end{aligned}
\end{equation}
where we omit the arguments $(\omega,\bm{r})$ for brevity. We call Eq.~\eqref{eq:compound_reciprocity_integral} the compound generalized reciprocity theorem. It relates a pair of sources and fields in two mutually compound transposed systems. When a system is compound reciprocal, Eq.~\eqref{eq:compound_reciprocity_integral} reduces to the symmetry constraint that relates a pair of sources and responses in that single system.

\item \emph{Original and compound $c$-transposed systems:} The compound $c$-transposed system [Eq.~\eqref{eq:Maxwell_compound_c-transpose}] is the $c$-transpose of the auxiliary system, so they are related by the modified mutual energy theorem [Eq.~(48) in Ref.~\cite{guo2022internal}]:
    \begin{equation}\label{eq:compound_mutual_energy_integral}
    \begin{aligned}[b]
         &-\oint_S (\bm{E}_O\times\bm{H}'^{*}_2+\bm{E}'^{*}_2\times\bm{H}_O) \cdot \bm{\mathrm{d}S} \\&= \int_V (\bm{E}'^*_2\cdot\bm{J}_O+\bm{E}_O\cdot\bm{J}'^*_2) \,\mathrm{d}V. 
    \end{aligned} 
\end{equation}
We call Eq.~\eqref{eq:compound_mutual_energy_integral} the compound modified mutual energy theorem. It relates a pair of sources and fields in two mutually compound $c$-transposed systems. When a system is compound energy-conserving, Eq.~\eqref{eq:compound_mutual_energy_integral} reduces to the symmetry constraint that relates a pair of sources and responses in that single system.

\item \emph{Original and compound conjugated systems:} The compound conjugated system [Eq.~\eqref{eq:Maxwell_compound_conjugate}] is the conjugate of the auxiliary system, so they are related by the generalized time-reversal theorem [Eq.~(50) in Ref.~\cite{guo2022internal}]:
\begin{equation}\label{eq:generalized_time_reversal}
\begin{aligned}[b]
   &M_{O}(\omega,\bm{r})\, \Phi_{O}(\omega,\bm{r}) = a_{O}(\omega,\bm{r}) \\&\iff M^*_O(\omega,\bm{r}) \, \Phi^*_O(\omega,\bm{r})  = a^*_O(\omega,\bm{r}),   
\end{aligned}
\end{equation}
where 
\begin{equation}
    \Phi^*_O(\omega,\bm{r}) = \begin{pmatrix}
    \bm{E}^*_O(\omega,\bm{r})  \\ -i\bm{H}^*_O(\omega,\bm{r}) 
    \end{pmatrix}, \;
    a^*_O(\omega,\bm{r}) =\begin{pmatrix}
    -i\bm{J}^*_O(\omega,\bm{r})  \\ 0
    \end{pmatrix}.
\end{equation}

Combining Eq.~\eqref{eq:generalized_time_reversal} with Eq.~\eqref{eq:external_symmetry_short}, we obtain
\begin{equation}\label{eq:compound_time_reversal}
\begin{aligned}[b]
    &M(\omega,\bm{r})\, \Phi(\omega,\bm{r}) = a(\omega,\bm{r}) \\&\iff M^*_O(\omega,\bm{r}) \, \Phi^*_O(\omega,\bm{r})  = a^*_O(\omega,\bm{r}).  
\end{aligned}
\end{equation}
We call Eq.~\eqref{eq:compound_time_reversal} the compound generalized time-reversal theorem. It states that $(\bm{E},\bm{H},\bm{J})$ satisfy Maxwell's equations for the original system if and only if $(\bm{E}^*_O,-\bm{H}^*_O,-\bm{J}^*_O)$ satisfy Maxwell's equations for the compound conjugated system. When the system is compound time-reversal symmetric, Eq.~\eqref{eq:compound_time_reversal} reduces to the symmetry constraint for that single system.
\end{enumerate}

\section{Physical implications for scattering matrices}\label{sec:S_matrix}

We have introduced external and compound transformations, defined the corresponding transformed systems, and derived the fundamental relations of electromagnetic fields between the original and transformed systems. We now discuss their implications for scattering matrices.

\subsection{Definition of scattering matrices}\label{subsec:definition_S_matrix}
\begin{figure}
    \centering
    \includegraphics[width=0.7\linewidth]{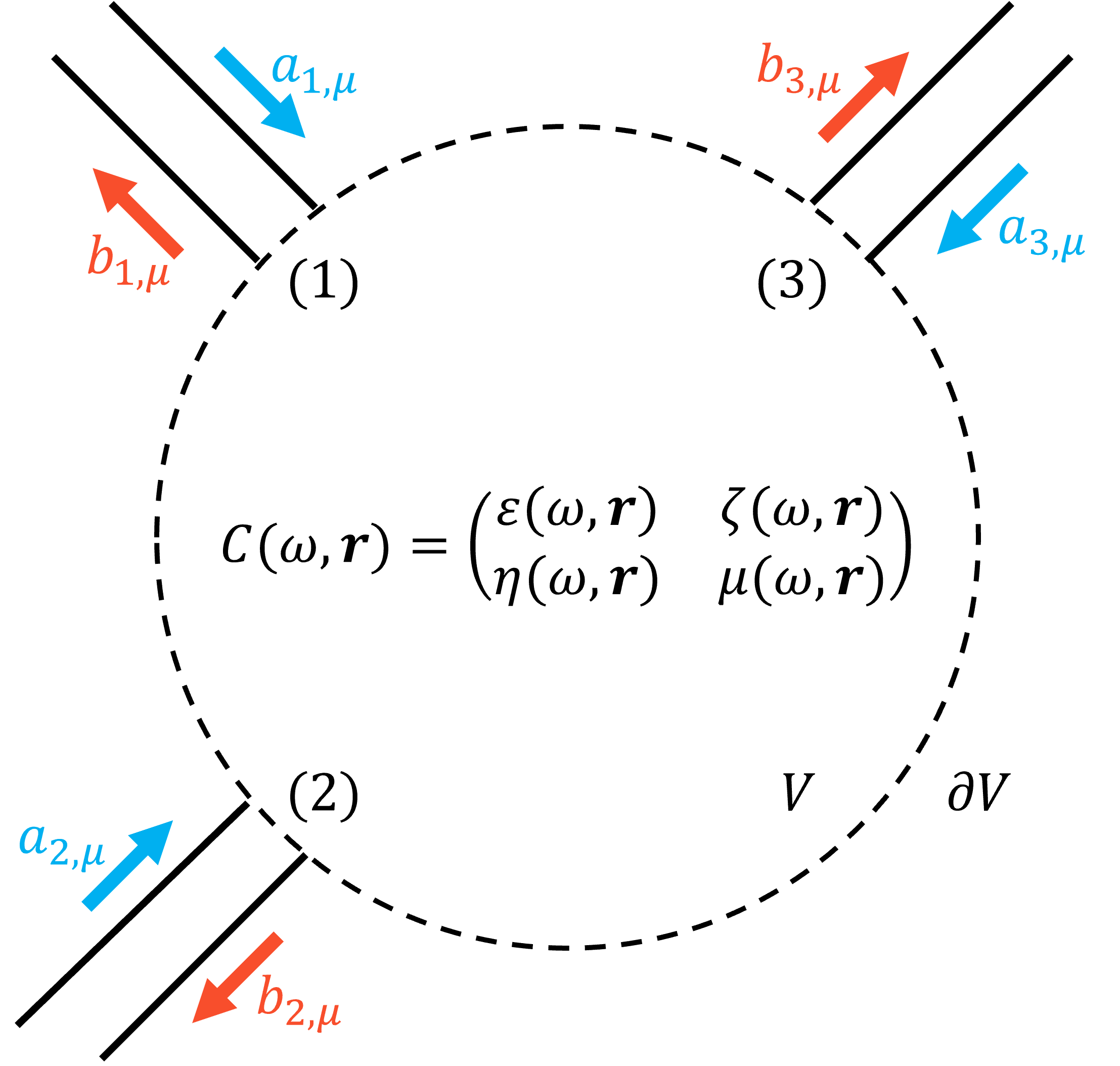}
    \caption{A general optical circuit. A linear time-invariant system characterized by $C(\omega,\bm{r})$ within a volume $V$ enclosed by a surface $\partial V$ is connected to its exterior by $Q$ physical ports ($Q=3$ shown). Each port is a waveguide made of a linear time-invariant medium that is homogeneous along the propagation direction, reciprocal, lossless, time-reversal symmetric, and reflection symmetric. Each port may support multiple modes, labeled by $\mu$. The $Q$ ports together support $P$ incoming and $P$ outgoing modes.}
    \label{fig:S_matrix}
\end{figure}

We briefly review the scattering matrix formalism, following the notation and conventions of Ref.~\cite{guo2022internal}; see also Refs.~\cite{haus1984,jalas2013,guo2023singular,guo2024j,li2025} for additional details.

As shown in Fig.~\ref{fig:S_matrix}, we consider a general source-free linear time-invariant system characterized by $C(\omega,\bm{r})$ within a volume $V$ enclosed by a surface $\partial V$. The system is connected to its exterior by $Q$ physical ports, which are linear time-invariant, homogeneous along the propagation direction, reciprocal, lossless, time-reversal symmetric, and reflection symmetric. The $Q$ ports support $P$ incoming and $P$ outgoing modes in total. We assume that the field energy outside $\partial V$ is carried exclusively by these guided modes, and that different physical ports and their guided modes are essentially nonoverlapping. In each physical port, we specify a Cartesian coordinate system such that $\bm{r}^t = (x,y)$ are tangential to $\partial V$, and $z$ is along the outgoing direction with $z=0$ at $\partial V$. Hereafter, we use the superscript $t$ to denote the transverse components of a vector field tangential to $\partial V$.

Under these assumptions, one can construct an orthonormal basis $\{\bm{e}_i, \bm{h}_i\}$ with purely real transverse fields $\{\bm{e}^t_i, \bm{h}^t_i\}$~\cite{jalas2013,mcisaac1991}, satisfying the orthonormality conditions
\begin{equation}\label{eq:waveguide_orthonormal_first}
\oint_{\partial V}  \bm{\mathrm{d} S} \cdot \bm{e}^t_i \times \bm{h}^t_j = \oint_{\partial V}  \bm{\mathrm{d} S} \cdot \bm{e}^t_j \times \bm{h}^t_i  = -2\, \delta_{ij},    \!\quad\! 1\leq i,j \leq P.
\end{equation}
The transverse fields in the physical ports outside $\partial V$ take the form
\begin{equation}
\label{eq:E_superposition_first}
    \bm{E}^t(\bm{r}^t,z) = \sum_{i=1}^P \left(a_i  e^{-i\beta_i z}  +  b_i e^{i\beta_i z}\right) \bm{e}^t_i(\bm{r}^t),
\end{equation}
\begin{equation}
\label{eq:H_superposition_first}
    \bm{H}^t(\bm{r}^t,z) = \sum_{i=1}^P \left(a_i e^{-i\beta_i z}  -   b_i e^{i\beta_i z}\right) \bm{h}^t_i(\bm{r}^t). 
\end{equation}
The incoming and outgoing waves are represented by complex vectors
\begin{equation}
    \bm{a} = (a_1,\, \ldots,\, a_P )^T, \qquad
    \bm{b} = (b_1,\, \ldots,\, b_P )^T,
\end{equation}
where $a_i$ and $b_i$ are the complex coefficients of the $i$th incoming and outgoing modes, respectively. Since the system is linear time-invariant and source-free, the outgoing waves are completely determined by the incoming waves and by the system properties. There is thus a linear relation between $\bm{a}$ and $\bm{b}$, which can be written in matrix form as
\begin{equation}\label{eq:S}
    \bm{b} = S \bm{a}. 
\end{equation}
The $P\times P$ matrix $S$ is the scattering matrix, with element $S_{ij}$ giving the scattering amplitude from the $j$th into the $i$th basis mode.

\subsection{Change of basis for a scattering matrix}\label{subsec:Dirac_notation}

We now derive the change of basis for a scattering matrix. For this purpose, it is useful to introduce Dirac notation, in which the ket $\ket{\varphi}$ denotes the transverse fields on the surface $\partial V$. For a general electromagnetic field,
\begin{equation}\label{eq:def_ket}
    \ket{\varphi} = \sum_i \left( a_i \ket{\phi_i^{(i)}} + b_i \ket{\phi_i^{(o)}} \right),
\end{equation}
where $a_i$ and $b_i$ are the incoming and outgoing modal amplitudes, respectively, and
\begin{equation}
    \ket{\phi_i^{(i)}} = \begin{pmatrix}
    \bm{e}_i^t(\bm{r}^t) \\ 
    \bm{h}_i^t(\bm{r}^t)
    \end{pmatrix},\! \quad \!
    \ket{\phi_i^{(o)}} = \begin{pmatrix}
    \bm{e}_i^t(\bm{r}^t) \\ 
    -\bm{h}_i^t(\bm{r}^t)
    \end{pmatrix},
    \!\quad\! i = 1,\ldots, P,
 \end{equation}
are the transverse modal profiles of the incoming and outgoing modes. We define the dual vectors $\bra{\phi_i^{(i)}}$ and $\bra{\phi_i^{(o)}}$ such that
\begin{equation}\label{eq:Dirac_orthonormal}
    \begin{aligned}[b]
       \braket{\phi_i^{(i)}|\phi_j^{(i)}} &= -\delta_{ij}, & \quad
       \braket{\phi_i^{(o)}|\phi_j^{(o)}} &= \delta_{ij}, \\
       \braket{\phi_i^{(i)}|\phi_j^{(o)}} &= 0, & \quad
       \braket{\phi_i^{(o)}|\phi_j^{(i)}} &= 0.
    \end{aligned}
\end{equation}
More details on the dual vectors are given in Appendix~\ref{app:Dirac_notation}.

The scattering properties of a linear photonic system are characterized by a linear scattering operator $\hat{S}$, which maps an incoming wave to the corresponding outgoing wave. This operator depends on the system $C(\omega,\bm{r})$ and the ports $C^{(m)}(\omega,\bm{r})$, $m=1,\ldots,Q$, but is independent of the modal basis. In the basis $\{\ket{\phi_i^{(i)}} ,\ket{\phi_i^{(o)}}\}$, $\hat{S}$ is represented as
\begin{equation}
\hat{S} = -\sum_{i,j} S_{ij} \ket{\phi_i^{(o)}}\bra{\phi_j^{(i)}},
\end{equation}
where the complex numbers $S_{ij}$ form the scattering matrix $S$. Using Eq.~\eqref{eq:Dirac_orthonormal}, we obtain
\begin{equation}\label{eq:Dirac_S_matrix}
    S_{ij} =  \braket{\phi_i^{(o)}|\hat{S}|\phi_j^{(i)}}.
\end{equation}

Unlike the scattering operator $\hat{S}$, the scattering matrix depends on the choice of modal basis. In another basis $\{\ket{\tilde{\phi}_i^{(i)}} ,\ket{\tilde{\phi}_i^{(o)}}\}$, the same operator $\hat{S}$ is represented by
\begin{equation}
\label{eq:Dirac_S_matrix_USUT}
\tilde{S} = USU^T = USU^{-1},
\end{equation}
where $U$ is the real orthogonal matrix defined by
\begin{equation}
U_{ij} \coloneqq
\braket{\tilde{\phi}_i^{(o)}|\phi_j^{(o)}}
=
-\braket{\phi_j^{(i)}|\tilde{\phi}_i^{(i)}} .
\end{equation}
The derivation of Eq.~\eqref{eq:Dirac_S_matrix_USUT} is provided in Appendix~\ref{app:Dirac_S_matrix_USUT}.

\subsection{Scattering matrices of transformed systems}\label{subsec:S_matrix_transformed_systems}

Applying external or compound transformations to the original optical circuit $\{C(\omega,\bm{r}),C^{(m)}(\omega,\bm{r})\}$ (equipped with the original mode basis $\{\ket{\phi_i^{(i)}} ,\ket{\phi_i^{(o)}}\}$) produces transformed optical circuits, each with its own scattering operator and matrix, which we now define.

For an external transformation $\mathcal{O}\in \mathrm{O}(3)$, the transformed circuit $\{C_O(\omega,\bm{r}),C_O^{(m)}(\omega,\bm{r})\}$ [Eq.~\eqref{eq:O_unified}] has a scattering operator $\hat{S}_O$. Under an arbitrary mode basis $\{\ket{\psi_i^{(i)}} ,\ket{\psi_i^{(o)}}\}$, $\hat{S}_O$ is represented by a scattering matrix $S_O$ with $S_{O;ij} =  \braket{\psi_i^{(o)}|\hat{S}_O|\psi_j^{(i)}}$.

For compound transformations, we further apply the three internal transformations to the system (but not the ports) of the externally transformed circuit, obtaining three transformed circuits $\{C_{k,O}(\omega,\bm{r}),C_{O}^{(m)}(\omega,\bm{r})\}$, $k=1,2,3$ (Sec.~\ref{subsec:three_class_compound}). Their scattering operators $\hat{S}_{k,O}$ are represented by scattering matrices $S_{k,O}$ under arbitrary mode bases $\{\ket{\psi_{k,i}^{(i)}} ,\ket{\psi_{k,i}^{(o)}}\}$, with $S_{k,O;ij} =  \braket{\psi_{k,i}^{(o)}|\hat{S}_{k,O}|\psi_{k,j}^{(i)}}$.

Our goal is to establish the relations between the original scattering matrix $S$ and the transformed scattering matrices $S_O$ and $S_{k,O}$.

\subsection{Constraints on scattering matrices under external transformations}\label{subsec:external_S_matrix}

We derive the relation between $S$ and $S_O$. To this end, we introduce a specific mode basis $\{\ket{\phi_{O;i}^{(i)}} ,\ket{\phi_{O;i}^{(o)}}\}$ for the externally transformed circuit, referred to as the ``induced basis,'' obtained by applying $\mathcal{O}$ to the original mode basis $\{\ket{\phi_i^{(i)}} ,\ket{\phi_i^{(o)}}\}$. We denote the scattering matrix in the induced basis as $\tilde{S}_O$, which in general differs from $S_O$ in the arbitrary basis $\{\ket{\psi_i^{(i)}} ,\ket{\psi_i^{(o)}}\}$.

We first show that
\begin{equation}\label{eq:S=tilde_S_O}
    S = \tilde{S}_O.
\end{equation}
This is intuitive: since both the circuit and the basis are transformed by $\mathcal{O}$ simultaneously, the scattering matrix is unchanged. We give a rigorous proof as follows: Consider a solution of Maxwell's equations for the original circuit: $\{\bm{E}(\bm{r}),\bm{H}(\bm{r})\}$ with transverse fields at $\partial V$ denoted $\ket{\varphi}= (\bm{E}^t(\bm{r}^t), \bm{H}^t(\bm{r}^t))^T$. By Eqs.~\eqref{eq:def_ket} and~\eqref{eq:Dirac_orthonormal}, the expansion coefficients of $\ket{\varphi}$ in the original basis $\{\ket{\phi_i^{(i)}} ,\ket{\phi_i^{(o)}}\}$ are
\begin{equation}
    \begin{aligned}[b]
    a_i &= - \braket{\phi_i^{(i)}|\varphi} \\&= \frac{1}{4} \oint_{\partial V} \bm{\mathrm{d}S} \cdot \left(
\bm{h}_i^t\times \bm{E}^t -  \bm{e}_i^t\times \bm{H}^t  \right) \\&= \frac{1}{4} \oint_{\partial V} \bm{\mathrm{d}S} \cdot \left(
\bm{h}_i\times \bm{E} -  \bm{e}_i\times \bm{H}  \right), \label{eq:a_i_integral}
\end{aligned}
\end{equation}
\begin{equation}
    \begin{aligned}[b]
    b_i &= \braket{\phi_i^{(o)}|\varphi} \\ &= \frac{1}{4} \oint_{\partial V} \bm{\mathrm{d}S} \cdot \left(
\bm{h}_i^t\times \bm{E}^t +  \bm{e}_i^t\times \bm{H}^t \right) \\&= \frac{1}{4} \oint_{\partial V} \bm{\mathrm{d}S} \cdot \left(
\bm{h}_i\times \bm{E} +  \bm{e}_i\times \bm{H}  \right).  \label{eq:b_i_integral} 
\end{aligned}
\end{equation}
By Eq.~\eqref{eq:external_symmetry_short}, the transformed fields $\{\bm{E}_O(\bm{r}),\bm{H}_O(\bm{r})\}$ [Eqs.~\eqref{eq:fields_O_E} and~\eqref{eq:fields_O_H}] satisfy Maxwell's equations for the transformed circuit. We denote their transverse fields at $\partial V$ as $\ket{\varphi_O}= (\bm{E}_O^t(\bm{r}^t), \bm{H}_O^t(\bm{r}^t))^T$. The expansion coefficients of $\ket{\varphi_O}$ in the induced basis $\{\ket{\phi_{O;i}^{(i)}} ,\ket{\phi_{O;i}^{(o)}}\}$ are 
\begin{equation}
\begin{aligned}[b]
        a_{O;i} &= - \braket{\phi_{O;i}^{(i)}|\varphi_O} \\&= \frac{1}{4} \oint_{\partial V} \bm{\mathrm{d}S}_O \cdot \left(
        \bm{h}_{O;i}\times \bm{E}_O -  \bm{e}_{O;i}\times \bm{H}_O  \right), \label{eq:a_Oi_integral}
\end{aligned}
\end{equation}
\begin{equation}
    \begin{aligned}[b]
    b_{O;i} &= \braket{\phi_{O;i}^{(o)}|\varphi_O} \\ &= \frac{1}{4} \oint_{\partial V} \bm{\mathrm{d}S}_O \cdot \left(
    \bm{h}_{O;i}\times \bm{E}_O +  \bm{e}_{O;i}\times \bm{H}_O  \right). \label{eq:b_Oi_integral} 
\end{aligned}
\end{equation}
All quantities with subscript $O$ are obtained from the original quantities under $\mathcal{O}$. The integrands in Eqs.~\eqref{eq:a_i_integral}--\eqref{eq:b_Oi_integral}, such as $\bm{\mathrm{d}S} \cdot \left(
\bm{h}_i\times \bm{E} -  \bm{e}_i\times \bm{H}  \right)$, are true scalar fields over $\partial V$; their integrals over the entire surface are invariant under orthogonal transformations, so $a_i = a_{O;i}$ and $b_i = b_{O;i}$ for all $1\leq i\leq P$. In vector form,
\begin{equation}\label{eq:a_equal_aO}
    \bm{a} = \bm{a}_{O}, \quad \bm{b} = \bm{b}_{O}. 
\end{equation}
Since 
\begin{equation}
    \bm{b} = S\bm{a}, \quad \bm{b}_{O} = \tilde{S}_O\bm{a}_{O}, 
\end{equation}
Eq.~\eqref{eq:a_equal_aO} holds for arbitrary $\bm{a}$ if and only if 
\begin{equation}\label{eq:S_equal_SO}
    S = \tilde{S}_O.
\end{equation}
This completes the proof of Eq.~(\ref{eq:S=tilde_S_O}). 

The relation between $S_O$ and $\tilde{S}_O$ follows from the basis
transformation in Sec.~\ref{subsec:Dirac_notation}. Defining an orthogonal matrix $U$ with elements
\begin{equation}
   U_{ij} =  \braket{\psi_i^{(o)}|\phi_{O;j}^{(o)}} = -\braket{\phi_{O;j}^{(i)}|\psi_i^{(i)}},  
\end{equation}
we obtain 
\begin{equation}\label{eq:S_O_U}
    S_O = U \tilde{S}_O U^{-1}.  
\end{equation}
Combining Eqs.~\eqref{eq:S_equal_SO} and~\eqref{eq:S_O_U} gives the relation between $S$ and $S_O$:
\begin{equation}\label{eq:S_O_S}
    S_O = U S U^{-1}.  
\end{equation}

\subsection{Constraints on scattering matrices under compound transformations}\label{subsec:compound_S_matrix}

We now derive the relations between $S$ and $S_{k,O}$, $k=1,2,3$. Each compound transformed system is obtained by applying an internal transformation to the externally transformed system. Since the scattering matrix relations under both internal~\cite{guo2022internal} and external transformations [Eq.~\eqref{eq:S_O_S}] have been established, we combine them to obtain the compound relations:

\begin{enumerate}[wide, labelwidth=!, labelindent=0pt, label=(\arabic*)]
\item \emph{Original and compound transposed systems:} The compound transposed system is the transpose of the externally transformed system, so~\cite{guo2022internal}
\begin{equation}\label{eq:S_1O-S_O}
    S_{1,O} = S_O^T.
\end{equation}
Combining with Eq.~\eqref{eq:S_O_S} gives
\begin{equation}\label{eq:S_1O-S}
    S_{1,O} = US^TU^{-1}.
\end{equation}

\item \emph{Original and compound $c$-transposed systems:} The compound $c$-transposed system is the $c$-transpose of the externally transformed system, so~\cite{guo2022internal}
\begin{equation}\label{eq:S_2O-S_O}
    S_{2,O} = (S_O^\dagger)^{-1}.
\end{equation}
Combining with Eq.~\eqref{eq:S_O_S} gives
\begin{equation}\label{eq:S_2O-S}
    S_{2,O} = U(S^\dagger)^{-1}U^{-1}.
\end{equation}

\item \emph{Original and compound conjugated systems:} The compound conjugated system is the conjugate of the externally transformed system, so~\cite{guo2022internal}
\begin{equation}\label{eq:S_3O-S_O}
    S_{3,O} = (S_O^*)^{-1}.
\end{equation}
Combining with Eq.~\eqref{eq:S_O_S} gives
\begin{equation}\label{eq:S_3O-S}
    S_{3,O} = U(S^*)^{-1}U^{-1}.
\end{equation}

\end{enumerate}

Thus, the fundamental field relations under external and compound transformations give the corresponding relations among the scattering matrices. When a system is invariant under any of these transformations, these general relations reduce to symmetry constraints on the scattering matrix of that single system, summarized in Table~\ref{tab:symmetries_compound}.

\begin{table}[htbp]
    \caption{\label{tab:symmetries_compound} Compound symmetries and $S$-matrix constraints.}
    \begin{ruledtabular}
        \begin{tabular}{lll}
            Compound symmetry & $C$-matrix  & $S$-matrix\\
            \hline\vspace{-8pt}\\
            Compound reciprocity & $C = C_{1,O}$ &  $S = US^TU^{-1}$ \\
            Compound energy conservation & $C = C_{2,O}$ & $S = U(S^\dagger)^{-1}U^{-1}$ \\
            Compound time-reversal~symmetry & $C = C_{3,O}$ & $S = U(S^*)^{-1}U^{-1}$ \\
        \end{tabular}
    \end{ruledtabular}
\end{table}

\section{Double antisymmetry groups}
\label{sec:antisymmetry_group_theory}

\subsection{Unified group theory of internal, external, and compound symmetries}

We now develop a unified group theory that encompasses internal, external, and compound symmetries.

The three internal transformations, together with the identity, form the Klein four-group $V_4 = \mathbb{Z}_{2} \times \mathbb{Z}_2$~\cite{guo2022internal}. The set of all external transformations constitutes the spatial symmetry group. For finite systems, this is the orthogonal group $\mathrm{O}(3)$, comprising all rotations and reflections. For infinite systems, it is the Euclidean group $\mathrm{E}(3)$, which additionally includes translations. We focus on $\mathrm{O}(3)$ below; the analysis extends readily to $\mathrm{E}(3)$.

Since any internal transformation commutes with any external transformation (Sec.~\ref{subsec:commutative}), the set of all internal, external, and compound transformations forms the direct product
\begin{equation}
    V_{4} \times \mathrm{O}(3) = \mathbb{Z}_{2} \times \mathbb{Z}_2 \times \mathrm{O}(3).
\end{equation}
The symmetry properties of any finite photonic structure correspond to a subgroup of $V_{4} \times \mathrm{O}(3)$. Classifying photonic systems thus reduces to enumerating all subgroups of $V_{4} \times \mathrm{O}(3)$, a problem solved by the theory of double antisymmetry groups.

\subsection{Antisymmetry groups}

We review the mathematical framework of double antisymmetry groups, starting with the more familiar single antisymmetry groups. A single antisymmetry group is constructed by adjoining a single anti-identity operation $\tilde{1}$ to the spatial group $\mathrm{O}(3)$. An anti-identity satisfies three properties~\cite{padmanabhan2020antisymmetry}:

\begin{enumerate}[label=(\arabic*)]
  \item $\tilde{1}$ is not an element of $\mathrm{O}(3)$.
  \item $\tilde{1}$ commutes with every element of $\mathrm{O}(3)$.
  \item $\tilde{1}\cdot\tilde{1}=1$, where $1$ denotes the identity.
\end{enumerate}

\noindent The set of all transformations---including the anti-identity, geometric transformations, and their products---forms the group
\begin{equation}
    \mathbb{Z}_2 \times \mathrm{O}(3).
\end{equation}
Its subgroups are the single antisymmetry groups, also known as Shubnikov groups or magnetic groups. These groups arise in many physical contexts, with the anti-identity representing different physical transformations. In magnetism, single antisymmetry groups characterize the symmetries of magnetic crystals, where the anti-identity is time reversal, which flips the magnetic moments. They also arise in thermal photonics, where the anti-identity is the reciprocity transformation that maps a system to its transposed system, generalizing the conventional Kirchhoff's law of thermal radiation to nonreciprocal objects~\cite{guo2022adjoint}.

Single antisymmetry groups generalize to double antisymmetry groups upon adjoining a second anti-identity $1'$ that commutes with $\tilde{1}$. The two anti-identities generate the group $\mathbb{Z}_2 \times \mathbb{Z}_2 = \{1, \tilde{1}, 1', \tilde{1}'\}$. The product $\tilde{1}' = \tilde{1} \cdot 1'$ is itself an anti-identity that commutes with both $\tilde{1}$ and $1'$; any two of the three anti-identities can serve as generators. The set of all transformations---including the three anti-identities, geometric transformations, and their products---forms the group
\begin{equation}
    \mathbb{Z}_2 \times \mathbb{Z}_2 \times \mathrm{O}(3).
\end{equation}
Its subgroups are the double antisymmetry groups.

In our context, the two anti-identity generators are the time-reversal operation $1^\ast$ and the reciprocity operation $1^T$. Both lie outside the spatial group $\mathrm{O}(3)$, commute with all spatial operations, and satisfy $(1^\ast)^2=(1^T)^2=1$. Their product $1^\dagger = 1^\ast 1^T$ is the third anti-identity, corresponding to energy conservation. Double antisymmetry groups therefore characterize the internal, external, and compound symmetry properties of photonic systems.

\subsection{Classification of photonic systems}

\begin{table*}[htbp]
    \caption{\label{Tab:DSG-space_group}Classification of double antisymmetry groups and counts of crystallographic double antisymmetry point groups (DAPGs) and double antisymmetry space groups (DASGs). Data from Refs.~\cite{vanleeuwen2014double,huang2013tables,litvin1994non,litvin1995magnetic}.}
    \setlength{\tabcolsep}{8pt}
    \renewcommand{\arraystretch}{1.0}
    \begin{NiceTabular}{C{0.5cm} C{10.1cm} C{1.9cm} C{1.8cm}} 
    \toprule
    Cat. & Symbol and Structure & DAPG\# & DASG\# \\
    \midrule
    (1)  & $\bm{G} = G$ & 32 & 230 \\
    (2)  & $\bm{G\onep} = G + G\onep$ & 32 & 230 \\
    (3)  & $\bm{G\oneps} = G + G\oneps$ & 32 & 230 \\
    (4)  & $\bm{G\ones} = G + G\ones$ & 32 & 230 \\
    (5)  & $\bm{G\onep\ones} = G + G\onep + G\ones + G\oneps$ & 32 & 230 \\
    (6)  & $\bm{G(H)} = H + (G-H)\onep$ & 58 & 1191 \\
    (7)  & $\bm{G(H)\{H\}} = H + (G-H)\oneps$ & 58 & 1191 \\
    (8)  & $\bm{G\{H\}} = H + (G-H)\ones$ & 58 & 1191 \\
    (9)  & $\bm{G(H)\oneps} = H + (G-H)\onep + H\oneps + (G-H)\ones$ & 58 & 1191 \\
    (10) & $\bm{G(H)\ones} = H + (G-H)\onep + H\ones + (G-H)\oneps$ & 58 & 1191 \\
    (11) & $\bm{G\{H\}\onep} = H + (G-H)\ones + H\onep + (G-H)\oneps$ & 58 & 1191 \\
    (12) & $\bm{G(H)\{K\}} = H \cap K + (H-K)\ones + (K-H)\onep + [G-(H+K)]\oneps$ & 116 & 9507 \\
    \midrule
    \multicolumn{2}{r}{Total:} & 624 & 17\,803 \\
    \midrule
    \multicolumn{4}{l}{\quad ``$-$'' denotes relative complement; ``$+$'' denotes union. $H, K$ are index-2 subgroups of $G$, i.e., $|G/H|=|G/K|=2$.} \\
    \bottomrule
    \end{NiceTabular}
\end{table*}

\begin{figure}[htbp]
    \centering
    \includegraphics[width=0.95\linewidth]{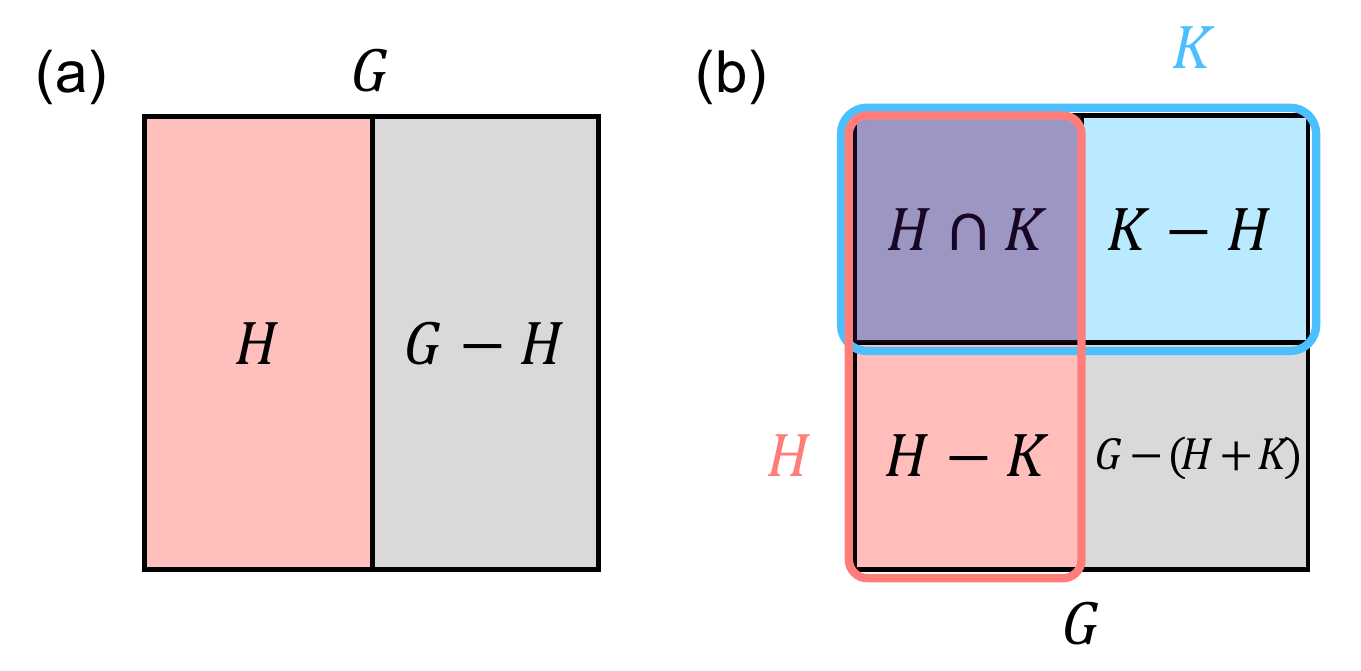}
    \caption{Partition of the parent group $G$ used to generate double antisymmetry groups. (a)~Partition into two halves: an index-2 subgroup $H$ and its coset $G - H$, used to generate categories~(6)--(11). (b)~Partition into four quarters by two index-2 subgroups $H$ and $K$, used to generate category~(12).}
    \label{fig:subgroup}
\end{figure}

We now classify all linear time-invariant photonic systems into twelve categories of double antisymmetry groups, as summarized in Table~\ref{Tab:DSG-space_group}. Each category starts from a parent spatial group $G$ and combines some or all of its elements with some or all of the three anti-identities $\onep$, $\oneps$, and $\ones$.

In categories (1)--(5), every element of $G$ is an external symmetry. Category~(1) has external symmetries only, represented by $G$ itself. Categories~(2)--(4) each adjoin one anti-identity to $G$, forming the direct product groups $\bm{G\onep}$, $\bm{G\oneps}$, and $\bm{G\ones}$. These represent systems that possess one internal symmetry---reciprocity, energy conservation, or time-reversal symmetry, respectively---and the compound symmetry generated by each element of $G$ paired with that anti-identity. In particular, all reciprocal lossy systems belong to category~(2). Category~(5) adjoins all three anti-identities, forming 
\begin{equation}
\bm{G\onep\ones} \coloneqq G + G\onep + G\ones + G\oneps.  
\end{equation}
This represents systems with all three internal symmetries; that is, reciprocal lossless systems.

In categories (6)--(8), only half of the elements of $G$ are external symmetries---they form an index-2 subgroup $H$ of $G$, as illustrated in Fig.~\ref{fig:subgroup}(a). The remaining elements in the coset $G - H$ are not external symmetries on their own but become compound symmetries when paired with an anti-identity. No internal symmetries are present. Following the notation of Refs.~\cite{litvin1994non,vanleeuwen2014double}, the three categories are 
\begin{align}
 \bm{G(H)} &\coloneqq H + (G-H)\onep, \\
 \bm{G(H)\{H\}} &\coloneqq H + (G-H)\oneps, \\
 \bm{G\{H\}} &\coloneqq H + (G-H)\ones.
\end{align}
These represent systems without any internal symmetry that possess exactly one type of compound symmetry---compound reciprocity, compound energy conservation, and compound time-reversal symmetry, respectively.

Categories (9)--(11) are obtained by adjoining an anti-identity to categories (6)--(8), respectively. For example, category~(9) is
\begin{equation}
\bm{G(H)\oneps} = H + (G-H)\onep + H\oneps + (G-H)\ones.    
\end{equation}
This represents systems that are energy-conserving ($\oneps$) and possess external symmetries $H$, compound energy-conserving symmetries $H\oneps$, compound reciprocal symmetries $(G-H)\onep$, and compound time-reversal symmetries $(G-H)\ones$. Categories~(10) and~(11) have analogous structures with internal symmetries $\ones$ and $\onep$, respectively (see Table~\ref{Tab:DSG-space_group} for details).

Category~(12) is the most complicated. We again start from a parent group $G$, but now consider two distinct index-2 subgroups $H$ and $K$ of $G$. One can show that $H \cap K$ is an index-4 subgroup of $G$, so $G$ is partitioned into four disjoint subsets, as illustrated in Fig.~\ref{fig:subgroup}(b):
\begin{equation}
    G = H \cap K + (H - K) + (K - H) + [G - (H + K)],
\end{equation}
where ``$+$'' denotes union and ``$-$'' denotes the relative complement. Category~(12) is obtained by adjoining each subset with a distinct decoration---$1$, $\ones$, $\onep$, and $\oneps$, respectively:
\begin{align}
    \bm{G(H)\{K\}} = &H \cap K + (H - K)\ones \nonumber \\ &+ (K - H)\onep + [G - (H + K)]\oneps.
\end{align}
This represents systems that possess all three types of compound symmetries but no internal symmetry. For a given $G$, there can be many choices of subgroup pairs $(H, K)$, making this category the richest in variations.

This twelve-category classification applies to all linear time-invariant photonic systems, whether finite, periodic, or aperiodic. The parent group $G$ captures the distinction: for finite systems, $G$ is a point group that may be crystallographic or noncrystallographic (such as the icosahedral or pentagonal groups); for periodic systems, $G$ is a space group. For three-dimensional crystallographic systems, the classification expands the standard 32 point groups and 230 space groups into 624 double antisymmetry point groups (DAPGs) and 17\,803 double antisymmetry space groups (DASGs). Categories~(1)--(5) each contribute 32 DAPGs and 230 DASGs. Categories~(6)--(11) each contribute 58 DAPGs and 1191 DASGs, matching the counts of the black-white magnetic groups~\cite{litvin1994non,litvin1995magnetic,litvin2001magnetic,souvignier2006four}. Category~(12) contributes 116 DAPGs and 9507 DASGs, giving the totals shown in Table~\ref{Tab:DSG-space_group}.

\begin{figure}[htbp]
    \centering
     \includegraphics[width=\linewidth]{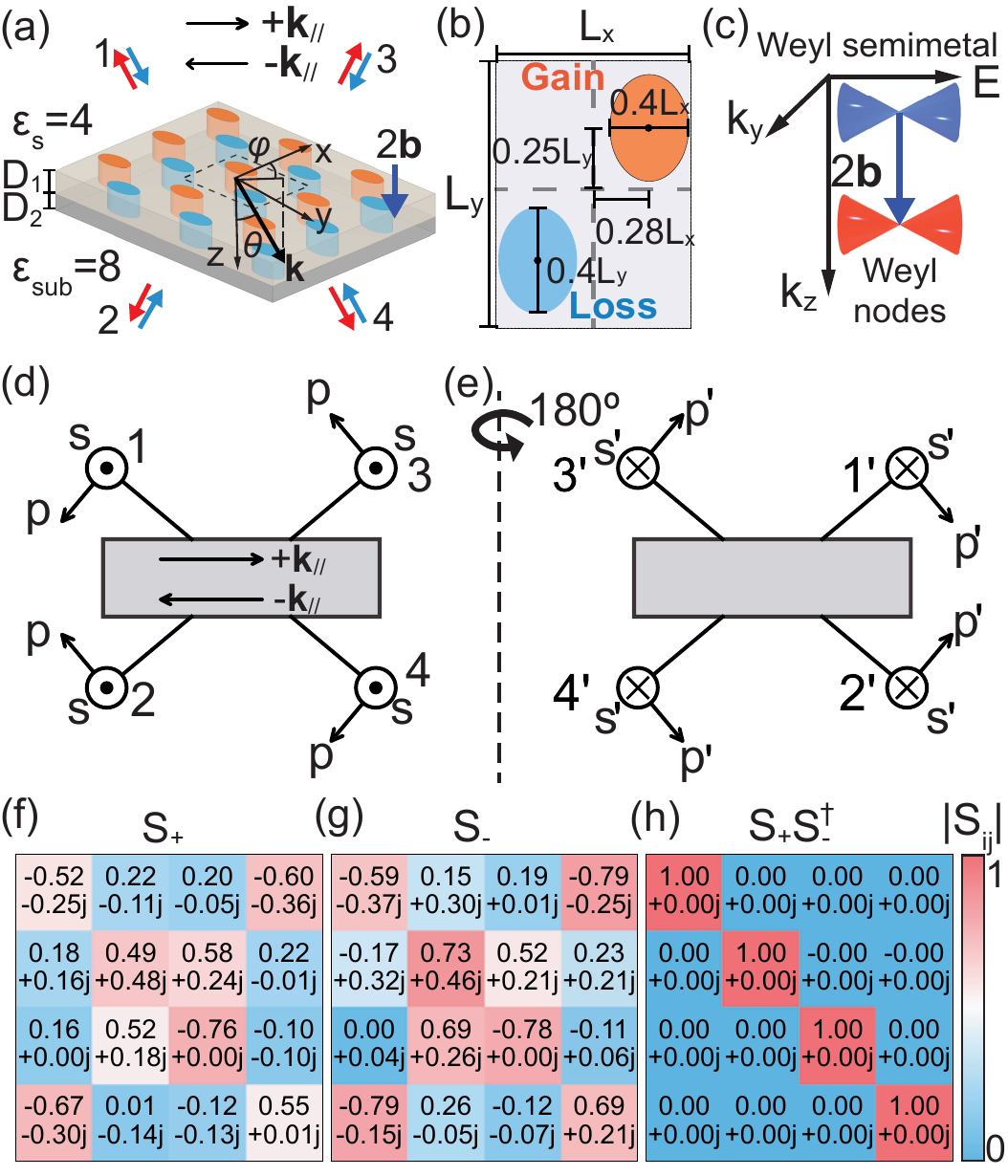}
     \caption{
A category~(7) photonic system with compound energy-conservation symmetry $C_{2z}1^\dagger$.
(a)~Photonic crystal slab ($\varepsilon_{\text{s}}=4$) on a homogeneous substrate ($\varepsilon_{\text{sub}}=8$), with lossy (blue) and gain (orange) elliptical magnetic Weyl-semimetal rods; $\pm\bm{k}_{\parallel}$ label the forward and reverse scattering processes described by the submatrices $S_\pm$.
(b)~Unit-cell geometry in the $xy$ plane.
(c)~Weyl-node separation $2\bm{b}$ along $+z$ in momentum space.
(d)~Original eight-port basis.
(e)~Transformed basis under $C_{2z}$.
Amplitude heatmaps:
(f)~$S_+$;
(g)~$S_-$;
(h)~$S_+S_-^\dagger=I$, required by $C_{2z}1^\dagger$.}
\label{fig:4}
\end{figure}

\subsection{Numerical examples}\label{subsec:numerical_examples}

We illustrate the classification with two numerical examples. Compound time-reversal symmetry (e.g., $\mathcal{PT}$ symmetry) has been studied extensively, and compound reciprocity has also been demonstrated~\cite{guo2022adjoint,ozaktas2026reciprocity}. We focus on systems exhibiting compound energy conservation.

Both examples are photonic crystal slabs in a dielectric host with relative permittivity $\varepsilon_{\text{s}}=4$, containing a periodic array of elliptical rods made of magnetic Weyl semimetals. The lattice periods are $L_x = 3~\mu$m and $L_y = 4~\mu$m, and the unit-cell geometry is shown in Fig.~\ref{fig:4}(b). The absolute permittivity tensor of the Weyl semimetal is~\cite{kotov2018giant,zhao2020,asadchy2020,guo2020r,guo2023light}
\begin{equation}
\varepsilon_\text{WSM}=\begin{pmatrix} \varepsilon_D & i\varepsilon_a & 0\\ -i\varepsilon_a & \varepsilon_D & 0\\ 0 & 0 & \varepsilon_D \end{pmatrix},
\end{equation}
where $\varepsilon_D$ is the Dirac permittivity [Eqs.~(26) and~(29) in Ref.~\cite{guo2023light}] and the nonreciprocal term
\begin{equation}
\varepsilon_{a} = \pm \frac{|\bm{b}| e^2}{2\pi^2\hbar\omega}
\end{equation}
is determined by the Weyl-node separation $2\bm{b}$. Here $2\bm{b}$ is an axial vector along $\pm z$ that acts as an effective magnetic field in momentum space [Fig.~\ref{fig:4}(c)] and transforms under spatial and time-reversal operations in the same way as a magnetic field. Magnetic Weyl semimetals can exhibit much stronger nonreciprocity than conventional magneto-optical media~\cite{kotov2018giant,guo2023light}. In our calculations, we choose the wavelength $\lambda = 5~\mu$m, with $\varepsilon_D/\varepsilon_0=2.9+0.3i$ in the lossy (blue) regions, $\varepsilon_D/\varepsilon_0=2.9-0.3i$ in the gain (orange) regions, and $\varepsilon_a/\varepsilon_0=\pm1.2$ for $2\bm{b}$ along $\pm z$, where $\varepsilon_0$ is the vacuum permittivity. These permittivity values are obtained from Eqs.~(5)--(7) of Ref.~\cite{kotov2018giant} using the following parameters: Fermi energy $E_F = 215$~meV, temperature $T = 300$~K, cutoff energy $E_c = 3 E_F$, Fermi velocity $v_F=1\times 10^6$~m/s, scattering time $\tau = 1$~ps, $N=24$ Weyl nodes, $|\bm{b}|=3.2\times 10^8~\text{m}^{-1}$, and $\varepsilon_\infty=13.2$.

We consider plane waves incident from four directions, labeled $1$ through $4$ in Fig.~\ref{fig:4}(a), with polar and azimuthal angles $(\theta,\varphi)=(50^\circ,30^\circ)$, $(130^\circ,30^\circ)$, $(50^\circ,210^\circ)$, and $(130^\circ,210^\circ)$, respectively. At this wavelength, only the zeroth diffraction order propagates. Including both polarizations in each direction, we obtain an eight-port system with the mode basis indicated in Fig.~\ref{fig:4}(d):
\begin{equation}
\mathcal{B}=\{\ket{1s},\ket{1p},\ket{2s},\ket{2p},\ket{3s},\ket{3p},\ket{4s},\ket{4p}\}.
\end{equation}
In this basis, the system is characterized by an $8\times8$ scattering matrix $S$. Because the in-plane wave vector $\bm{k}_\parallel$ is conserved, $S$ takes the block form
\begin{equation}\label{eq:example-S-block}
S = \begin{pmatrix} 0 & S_{-} \\ S_{+} & 0
\end{pmatrix},
\end{equation}
where the $4\times4$ submatrix $S_+$ describes forward ($+\bm{k}_\parallel$) scattering from ports $1$, $2$ to ports $3$, $4$, and $S_-$ describes the reverse ($-\bm{k}_\parallel$) process. We compute $S$ using rigorous coupled-wave analysis~\cite{Liu20122233} with 527 plane waves.  

\textit{Example~1: Category~(7).} The first structure is a photonic crystal slab of thickness $D_1 = 15~\mu$m on a uniform substrate of thickness $D_2 = 10.5~\mu$m and relative permittivity $\varepsilon_{\text{sub}}=8$ [Fig.~\ref{fig:4}(a)], with the Weyl-node separation $2\bm{b}$ along $+z$. Figures~\ref{fig:4}(f) and~\ref{fig:4}(g) show the computed submatrices $S_+$ and $S_-$, respectively.

This structure belongs to a double antisymmetry point group of category~(7),
\begin{equation}
    \bm{G(H)\{H\}}=\{1,\, C_{2z} 1^\dagger\},
\end{equation}
with parent spatial group $G=\{1,\, C_{2z}\}$ and subgroup $H = \{1\}$. The twofold rotation $C_{2z}$ preserves $2\bm{b}$ but spatially interchanges the loss and gain regions; $1^\dagger$ preserves $2\bm{b}$ and locally converts loss into gain and vice versa. Their composition $C_{2z}1^\dagger$ therefore leaves the system invariant, constituting a compound energy-conservation symmetry.

We derive the scattering-matrix constraint imposed by $C_{2z} 1^\dagger$. The $C_{2z}$ operation rotates the system by $180^\circ$ about the $z$ axis [Fig.~\ref{fig:4}(e)]:
\begin{equation}
1\leftrightarrow3, \quad 2\leftrightarrow 4, \quad s\mapsto-s, \quad p\mapsto -p,
\end{equation}
yielding the transformed basis
\begin{equation}
    \mathcal{B}'=\{\ket{1s}',\ket{1p}',\ket{2s}',\ket{2p}',\ket{3s}',\ket{3p}',\ket{4s}',\ket{4p}'\}.
\end{equation}
The scattering matrix of the rotated structure, expressed in the original basis $\mathcal{B}$, is
\begin{equation}
S_{C_{2z}} = U_1SU_1^{T},
\label{eq:120}
\end{equation}
where the orthogonal matrix $U_1$, defined by $\mathcal{B} = \mathcal{B'}U_1$, takes the form
\begin{equation}
U_1 = \begin{pmatrix}
0 & -I \\-I & 0
\end{pmatrix},
\end{equation}
with $I$ the $4\times 4$ identity matrix. Because $C_{2z}1^\dagger$ is a symmetry of the system, the rotated system is identical to the $c$-transposed system, giving
\begin{equation}
S_{C_{2z}} = (S^\dagger)^{-1}.
\label{eq:121}
\end{equation}
Combining Eqs.~\eqref{eq:example-S-block},~\eqref{eq:120}, and~\eqref{eq:121} gives the compound symmetry constraint on the submatrices,
\begin{equation}
S_+ S_-^\dagger = I,
\label{eq:1dagC2z}
\end{equation}
which is confirmed numerically in Fig.~\ref{fig:4}(h). By contrast, an energy-conserving system would satisfy
\begin{equation}
S_+ S_+^\dagger = S_- S_-^\dagger = I,
\label{eq:1dag}
\end{equation}
with each submatrix alone being unitary. Equation~\eqref{eq:1dagC2z} instead couples $S_+$ and $S_-$, illustrating the distinct physical implications of compound energy conservation.

\begin{figure}[htbp]
    \centering
    \includegraphics[width=\linewidth]{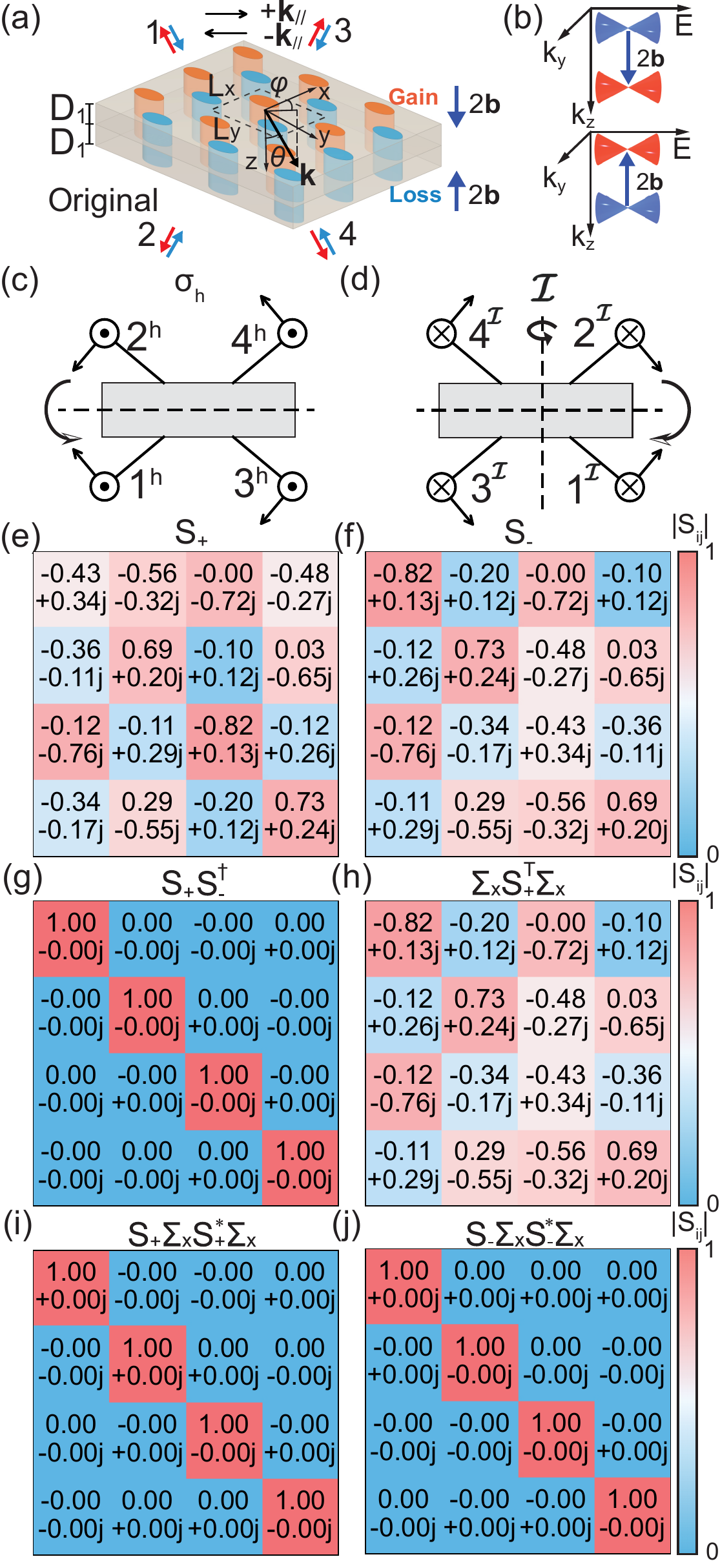}
    \caption{A category~(12) photonic system with compound symmetries $C_{2z}1^\dagger$, $\sigma_h1^T$, and $\mathcal{I}1^\ast$.
(a)~Bilayer photonic crystal slab with oppositely directed Weyl-node separations.
(b)~Weyl-node separations $2\bm{b}$ for the two layers.
Transformed eight-port basis under (c)~$\sigma_h$ and (d)~$\mathcal{I}$.
Amplitude heatmaps:
(e)~$S_+$;
(f)~$S_-$;
(g)~$S_+S_-^\dagger=I$, required by $C_{2z}1^\dagger$;
(h)~$\Sigma_x S_+^T \Sigma_x=S_-$, required by $\sigma_h1^T$;
(i)~$S_+\Sigma_x S_+^\ast \Sigma_x = I$ and (j)~$S_-\Sigma_x S_-^\ast \Sigma_x = I$, required by $\mathcal{I}1^\ast$.}
    \label{fig:5}    
\end{figure}

\textit{Example~2: Category~(12).} The second structure replaces the substrate with another photonic crystal slab of the same pattern as the top layer but with $2\bm{b}$ reversed to point along $-z$ [Figs.~\ref{fig:5}(a) and~\ref{fig:5}(b)]. Figures~\ref{fig:5}(e) and~\ref{fig:5}(f) show the computed submatrices $S_+$ and $S_-$, respectively.

This structure belongs to a double antisymmetry point group of category~(12),
\begin{equation}
\bm{G(\mathcal{H})\{\mathcal{K}\}}=\{1,\; C_{2z}1^\dagger,\; \sigma_h 1^T,\; \mathcal{I}\,1^\ast\},
\end{equation}
where ${\mathcal{H}}=\{1,\mathcal{I}\}$ and ${\mathcal{K}}=\{1,\sigma_h\}$. The compound symmetry $C_{2z} 1^\dagger$ is analogous to that in Example~1. The horizontal mirror $\sigma_h$ preserves the loss and gain distribution but reverses $2\bm{b}$; $1^T$ likewise preserves loss and gain and reverses $2\bm{b}$. Their composition $\sigma_h 1^T$ therefore leaves the system invariant, constituting a compound reciprocity symmetry. The product of $C_{2z} 1^\dagger$ and $\sigma_h 1^T$ gives the compound time-reversal symmetry $\mathcal{I}1^\ast$.

Equation~\eqref{eq:1dagC2z} provides the scattering-matrix constraint imposed by $C_{2z} 1^\dagger$ [Fig.~\ref{fig:5}(g)]. We now derive the constraint imposed by $\sigma_h1^T$. The $\sigma_h$ operation reflects the system across the horizontal plane [Fig.~\ref{fig:5}(c)]:
\begin{equation}
1\leftrightarrow2, \quad 3\leftrightarrow 4, \quad s\mapsto s, \quad p\mapsto p,
\end{equation}
yielding the transformed basis
\begin{equation}
\mathcal{B}^h\!=\!\{\ket{1s}^h\!,\ket{1p}^h\!,\ket{2s}^h\!,\ket{2p}^h\!,\ket{3s}^h\!,\ket{3p}^h\!,\ket{4s}^h\!,\ket{4p}^h\}.
\end{equation}
The scattering matrix of the mirrored structure, expressed in the original basis $\mathcal{B}$, is
\begin{equation}
S_{\sigma_h} = U_2SU_2^{T},
\label{eq:SV}
\end{equation}
where the orthogonal matrix $U_2$, defined by $\mathcal{B} = \mathcal{B}^h U_2$, takes the form
\begin{equation}
U_2 = \begin{pmatrix}
\Sigma_x & 0   \\0 & \Sigma_x
\end{pmatrix}, \quad \text{with} \quad  \Sigma_x=\begin{pmatrix} 0 & I_2\\ I_2 & 0 \end{pmatrix},
\end{equation}
and $I_2$ is the $2\times 2$ identity matrix. Because $\sigma_h 1^T$ is a symmetry of the system, the mirrored system is identical to the transposed system, giving
\begin{equation}
S_{\sigma_h} = S^T.
\label{eq:ST}
\end{equation}
Combining Eqs.~\eqref{eq:example-S-block},~\eqref{eq:SV}, and~\eqref{eq:ST} gives the compound symmetry constraint on the submatrices,
\begin{equation}
\Sigma_xS_+^T\Sigma_x=S_-,
\label{eq:1T}
\end{equation}
which is confirmed numerically in Figs.~\ref{fig:5}(f) and~\ref{fig:5}(h). By contrast, a reciprocal system would satisfy
\begin{equation}
S_+^T = S_-.
\label{eq:reciprocal_standard}
\end{equation}
Equation~\eqref{eq:1T} instead involves $\Sigma_x$, illustrating the distinct physical implications of compound reciprocity.

We similarly derive the constraint imposed by $\mathcal{I}1^\ast$. The $\mathcal{I}$ operation spatially inverts the system [Fig.~\ref{fig:5}(d)]:
\begin{equation}
1\leftrightarrow4, \quad 2\leftrightarrow 3, \quad s\mapsto -s, \quad p\mapsto -p,
\end{equation}
yielding the transformed basis
\begin{equation}
\mathcal{B}^\mathcal{I}\!=\!\{\ket{1s}^\mathcal{I}\!,\ket{1p}^\mathcal{I}\!,\ket{2s}^\mathcal{I}\!,\ket{2p}^\mathcal{I}\!,\ket{3s}^\mathcal{I}\!,\ket{3p}^\mathcal{I}\!,\ket{4s}^\mathcal{I}\!,\ket{4p}^\mathcal{I}\}.
\end{equation}
The scattering matrix of the inverted structure, expressed in the original basis $\mathcal{B}$, is
\begin{equation}\label{eq:S_I_1}
S_{\mathcal{I}}=U_3SU_3^{T},
\end{equation}
where the orthogonal matrix $U_3$, defined by $\mathcal{B} = \mathcal{B}^\mathcal{I} U_3$, takes the form
\begin{equation}
U_3 = \begin{pmatrix} 0 &  -\Sigma_x  \\ -\Sigma_x & 0
\end{pmatrix}.
\end{equation}
Because $\mathcal{I}1^\ast$ is a symmetry of the system, the inverted system is identical to the conjugated system, giving
\begin{equation}\label{eq:S_I_2}
S_{\mathcal{I}}=(S^\ast)^{-1}.
\end{equation}
Combining Eqs.~\eqref{eq:example-S-block},~\eqref{eq:S_I_1}, and~\eqref{eq:S_I_2} gives the compound symmetry constraints on the submatrices,
\begin{equation}\label{eq:1star}
S_+\Sigma_xS_+^\ast\Sigma_x =I \quad \text{and} \quad S_-\Sigma_xS_-^\ast\Sigma_x =I,
\end{equation}
which can equivalently be obtained by combining Eqs.~\eqref{eq:1dagC2z} and~\eqref{eq:1T}. Figures~\ref{fig:5}(i) and~\ref{fig:5}(j) confirm Eq.~\eqref{eq:1star} numerically. By contrast, a time-reversal-symmetric system would satisfy
\begin{equation}
S_+ S_-^\ast =I \quad \text{and} \quad S_-S_+^\ast =I.
\end{equation}
Equation~\eqref{eq:1star} instead constrains $S_+$ and $S_-$ independently, illustrating the distinct physical implications of compound time-reversal symmetry.

\subsection{Physical application}

We now apply our theory to an unsettled question in thermal photonics: whether a gyrotropic sphere obeys Kirchhoff's law of thermal radiation.

\begin{figure}
    \centering
    \includegraphics[width=1\linewidth]{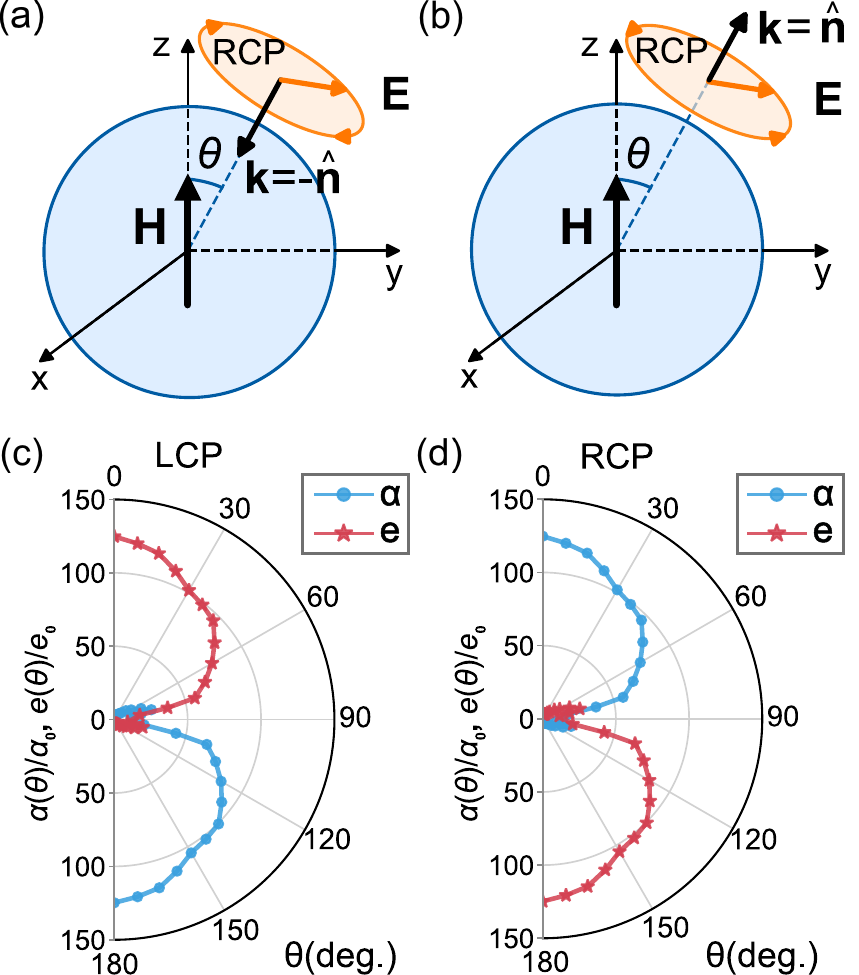}
        \caption{Absorptivity and emissivity of an InSb sphere in an external magnetic field $H$. Schematics of the scattering geometries for (a)~absorption and (b)~emission. Angular dependence of the absorptivity $\alpha$ and emissivity $e$ for (c)~left-circular polarization (LCP) and (d)~right-circular polarization (RCP). The values are normalized to the zero-field ($H=0$~T) value $\alpha_0=e_0$, which is independent of $\theta$. Rather than obeying Kirchhoff's law, absorptivity and emissivity are linked by $\alpha(\theta,\mathrm{RCP})=e(\theta,\mathrm{LCP})$.}
    \label{fig:6}
\end{figure}

Consider an InSb sphere of radius $R=0.5~\mu$m subjected to a magnetic field $H=5$~T along the $z$ direction [Fig.~\ref{fig:6}(a)]. At the wavelength $\lambda=33.5~\mu$m, InSb has the gyrotropic relative permittivity tensor
\begin{equation}
\varepsilon =
\begin{pmatrix}
4.89 + 1.77i & -1.65 - 6.69i & 0 \\
1.65 + 6.69i & 4.89 + 1.77i & 0 \\
0 & 0 & 9.56 + 0.31i
\end{pmatrix},
\label{eq:tensor}
\end{equation}
which is asymmetric and thus breaks reciprocity.

We compare the absorptivity $\alpha$ for plane waves incident along $-\hat{\mathbf{n}}$ [Fig.~\ref{fig:6}(a)] with the emissivity $e$ along $\hat{\mathbf{n}}$ [Fig.~\ref{fig:6}(b)]. For reciprocal objects, Kirchhoff's law of thermal radiation states that
\begin{equation}
\alpha(-\hat{\mathbf{n}},\hat{\mathbf{p}}) = e(\hat{\mathbf{n}},\hat{\mathbf{p}}^{*}),
\label{eq:Kirchhoff}
\end{equation}
where $\hat{\mathbf{p}}$ is the polarization vector and $\hat{\mathbf{p}}^{*}$ its complex conjugate. Since the gyrotropic response breaks reciprocity, Eq.~\eqref{eq:Kirchhoff} need not hold~\cite{zhu2014,
ben-abdallah2016,zhu2016,guo2019a}. Nonetheless, Ref.~\cite{abraham2017thermal} found that it does hold for linear polarizations and concluded that gyrotropic spheres obey Kirchhoff's law.

Kirchhoff's law, however, requires Eq.~\eqref{eq:Kirchhoff} to hold for every $\hat{\mathbf{p}}$; we therefore ask whether this conclusion extends beyond linear polarizations. Using full-wave simulations with the MESTI package~\cite{lin2022_APF,MESTI}, we compute $\alpha$ and $e$ for left- and right-handed circular polarizations (LCP and RCP) as functions of the polar angle $\theta$. For circular polarizations, Kirchhoff's law [Eq.~\eqref{eq:Kirchhoff}] reduces to
\begin{equation}
\alpha(\theta,\mathrm{LCP}) = e(\theta,\mathrm{LCP}),\quad \alpha(\theta,\mathrm{RCP}) = e(\theta,\mathrm{RCP}).
\label{eq:Kirchhoff-circular}
\end{equation}
Our results [Figs.~\ref{fig:6}(c) and~\ref{fig:6}(d)] show instead that
\begin{equation}
\alpha(\theta,\mathrm{LCP}) \neq e(\theta,\mathrm{LCP}),\quad \alpha(\theta,\mathrm{RCP}) \neq e(\theta,\mathrm{RCP}),
\label{eq:rcp-violation}
\end{equation}
so Kirchhoff's law is violated for circular polarizations. Absorptivity and emissivity are nevertheless linked by
\begin{equation}
\alpha(\theta,\mathrm{RCP}) = e(\theta,\mathrm{LCP}).
\label{eq:rcp-lcp}
\end{equation}
This relation can be explained by our symmetry analysis. The gyrotropic sphere belongs to category~(6) and possesses the compound reciprocity symmetry $\sigma_{v}1^{T}$, where $\sigma_{v}$ is the mirror reflection about the plane containing $\hat{\mathbf{n}}$ and the $z$ axis. Because $\sigma_{v}$ exchanges RCP and LCP, this symmetry, together with the adjoint Kirchhoff's law of thermal radiation for nonreciprocal objects~\cite{guo2022adjoint}, enforces Eq.~\eqref{eq:rcp-lcp} in place of Eq.~\eqref{eq:Kirchhoff-circular}. This example illustrates how our group classification reveals unconventional symmetry constraints in concrete photonic systems.

\section{Conclusion}
\label{sec:conclusion}

In this paper, we developed a unified theory of compound symmetries associated with reciprocity, energy conservation, and time reversal in linear photonic systems. We defined compound transformations as compositions of these three internal transformations with external (spatial) transformations, and identified three classes of compound symmetry: compound reciprocity, compound energy conservation, and compound time-reversal symmetry. We derived the fundamental relations of electromagnetic fields under these compound transformations and established the corresponding constraints on scattering matrices. We showed that internal, external, and compound symmetries are naturally described by double antisymmetry groups, giving a classification of linear time-invariant photonic systems into twelve symmetry categories, each imposing characteristic constraints on the electromagnetic response. We illustrated two representative categories with numerical examples and applied the theory to examine Kirchhoff's law of thermal radiation for a gyrotropic sphere.

Symmetry-based classification of photonic systems allows one to systematically determine the fundamental constraints for each class. Such investigations have already produced results, including the adjoint Kirchhoff's law for all thermal emitters~\cite{guo2022adjoint} and reciprocity-based compound symmetry constraints in chiral and Tellegen media~\cite{ozaktas2026reciprocity}. The framework developed here provides a systematic foundation for further exploration and engineering of symmetry in photonic systems.

Several extensions of the present framework are of immediate interest. The three internal transformations considered here---reciprocity, energy conservation, and time reversal---can be supplemented by additional internal symmetries, further enriching the classification. The group-theoretic construction developed here also extends naturally beyond photonics to other linear wave systems, including acoustic and elastic waves. Moreover, our analysis focuses on linear time-invariant systems; it would be interesting to extend the present framework to linear time-varying systems, such as moving media. Together, these directions offer a route toward a unified symmetry-based understanding of linear wave physics.

\begin{acknowledgments}
C.G. is supported by the Jack Kilby/Texas Instruments Endowed Faculty Fellowship. Y.W. and C.G. acknowledge support from the Texas Quantum Institute (TQI) at The University of Texas at Austin.
\end{acknowledgments}

\section*{Data availability}

The data that support the findings of this article are not publicly available upon publication because it is not technically feasible and/or the cost of preparing, depositing, and hosting the data would be prohibitive within the terms of this research project. The data are available from the authors upon reasonable request.

\appendix

\section{Proof of Eqs.~\eqref{eq:curl_theorem_v} and~\eqref{eq:curl_theorem_vbar}}\label{appendix:proof_curl}

We prove Eqs.~\eqref{eq:curl_theorem_v} and~\eqref{eq:curl_theorem_vbar}, reproduced here:
\begin{flalign}
\label{eq:curl_theorem_v_copy}
\bm{\nabla} \times [Ov (O^{-1}\bm{r})] = \det(O)\,O (\bm{\nabla} \times v) (O^{-1}\bm{r}),  \\
\label{eq:curl_theorem_vbar_copy} 
\bm{\nabla} \times [\det(O)\,O\bar{v} (O^{-1}\bm{r})] = O (\bm{\nabla} \times \bar{v}) (O^{-1}\bm{r}). 
\end{flalign}

We adopt the following conventions: As in the main text, the index $i=1,2,3$ denotes the $i$th component in the Cartesian basis $\{\bm{e}_i\}$, and the position vector is $\bm{r} =(x_1,x_2,x_3)^T$. All index summations run from $1$ to $3$ and are written explicitly rather than by Einstein convention. We further define the extended indices $0\equiv3$ and $4\equiv1$, which allow a compact form for the cross product of two complex vectors $\bm{a}$ and $\bm{b}$:
\begin{equation}\label{eq:cross_product}
    (\bm{a} \times \bm{b})_i = a_{i+1} b_{i-1} - a_{i-1} b_{i+1}, \quad i = 1,2,3.
\end{equation}

We first prove Eq.~\eqref{eq:curl_theorem_v_copy}. For a complex polar vector field $v(\bm{r})$, the curl $\bm{\nabla} \times v (\bm{r})$ has components
\begin{equation}
 [\bm{\nabla} \times v (\bm{r})]_i \,=\, \frac{\partial v_{i-1}(\bm{r})}{\partial x_{i+1}} - \frac{\partial v_{i+1}(\bm{r})}{\partial x_{i-1}}. 
\end{equation}
Under an orthogonal transformation $\mathcal{O}$, 
\begin{equation}
    v(\bm{r}) \to Ov(O^{-1}\bm{r}),
\end{equation}
and the corresponding curl has components
\begin{equation}\label{eq:nabla_Ov}
\{\bm{\nabla} \times [Ov (O^{-1}\bm{r}) ] \}_i = \frac{\partial [Ov]_{i-1}(O^{-1}\bm{r})}{\partial x_{i+1}} - \frac{\partial [Ov]_{i+1}(O^{-1}\bm{r})}{\partial x_{i-1}}.   
\end{equation}
Denoting $\bm{r'} = O^{-1}\bm{r} = (x_1', x_2', x_3')^T$, we simplify Eq.~\eqref{eq:nabla_Ov} as follows: For any vector field $u(\bm{r})$ and $m, n = 1, 2, 3$,  
\begin{equation}
    \begin{aligned}[b]
\frac{\partial u_m(O^{-1}\bm{r})}{\partial x_n} \, &= \,\sum_q \frac{\partial x'_q}{\partial x_n}\frac{\partial u_m (\bm{r'})}{\partial x'_q} \,\\&=\,\sum_q [O^{-1}]_{qn} \frac{\partial u_m(\bm{r'})}{\partial x'_q} \,\\&=\, \sum_q O_{nq} \frac{\partial u_m(\bm{r'})}{\partial x'_q},  
\end{aligned}
\end{equation}
where we used $O^{-1}=O^T$. Setting $u=Ov$ gives
\begin{equation}\label{eq:Ov_partial_mn}
    \begin{aligned}[b]
    \frac{\partial [Ov]_m(O^{-1}\bm{r})}{\partial x_n} \, &=\,\sum_{p, q} O_{nq} \frac{\partial O_{mp} v_p(\bm{r'})}{\partial x'_q} \,\\&=\, \sum_{p, q}  O_{nq} O_{mp} \frac{\partial  v_p(\bm{r'})}{\partial x'_q}. 
    \end{aligned}
\end{equation}
Substituting Eq.~\eqref{eq:Ov_partial_mn} into Eq.~\eqref{eq:nabla_Ov} gives
\begin{equation}\label{eq:nabla_OV_sum_pq}
\begin{aligned}[b]
    &\{\bm{\nabla} \times [Ov (O^{-1}\bm{r})]\}_i \\ 
& = \sum_{p,q}(O_{i+1,q}O_{i-1,p}-O_{i-1,q}O_{i+1,p}) \frac{\partial  v_p(\bm{r'})}{\partial x'_q}. 
\end{aligned}
\end{equation}

Since $O_{m,n}$ is the $m$th component of the new basis vector $\bm{e}'_n = \mathcal{O} \bm{e}_n$ in the original basis, $O_{m,n} = (\bm{e}'_n)_m$. Comparing with Eq.~\eqref{eq:cross_product} gives
\begin{equation}\label{eq:OO-OO}
O_{i+1,q}O_{i-1,p}-O_{i-1,q}O_{i+1,p} = (\bm{e}'_q\times \bm{e}'_p)_i.
\end{equation}
The cross product of the new basis vectors satisfies
\begin{equation}\label{eq:Levi-Civita}
  \bm{e}'_q\times \bm{e}'_p = \det(O) \,\sum_{k=1}^3\epsilon_{qpk}\, \bm{e}'_k, \qquad q,p,k = 1,2,3,
\end{equation}
where $\epsilon_{qpk}$ is the Levi-Civita symbol and $\det(O)$ accounts for the sign change when $\mathcal{O}$ is improper and $\{\bm{e}_i'\}$ becomes left-handed. Using the extended indices $0\equiv3$ and $4\equiv1$, Eq.~\eqref{eq:Levi-Civita} becomes
\begin{equation}\label{eq:Levi-Civita-cases}
  \bm{e}'_q\times \bm{e}'_p = 
  \begin{cases}
  0, & q=p \\
  \det(O)\, \bm{e}'_k, & q=k+1,\; p=k-1 \\
  -\det(O)\,\bm{e}'_k, & q=k-1,\; p=k+1
  \end{cases}
\end{equation}
with $k \in \{1,2,3\}$. One can verify Eq.~\eqref{eq:Levi-Civita-cases} is equivalent to Eq.~\eqref{eq:Levi-Civita}.

Substituting Eq.~\eqref{eq:OO-OO} into Eq.~\eqref{eq:nabla_OV_sum_pq} and using Eq.~\eqref{eq:Levi-Civita-cases}, we obtain
\begin{align}
&\{\bm{\nabla} \times [Ov (O^{-1}\bm{r})]\}_i \nonumber \\
&= \sum_{p,q}(\bm{e}'_q\times \bm{e}'_p)_i  \frac{\partial  v_p(\bm{r'})}{\partial x'_q}\nonumber\\
&= \sum_k \left[ (\bm{e}'_{k+1}\times \bm{e}'_{k-1})_i\frac{\partial  v_{k-1}(\bm{r'})}{\partial x'_{k+1}} \right. \nonumber\\
& \qquad  \quad \left. +  (\bm{e}'_{k-1}\times \bm{e}'_{k+1})_i \frac{\partial  v_{k+1}(\bm{r'})}{\partial x'_{k-1}} \right] \nonumber\\
&=  \sum_k \det(O) (\bm{e}'_{k})_i \left[\frac{\partial  v_{k-1}(\bm{r'})}{\partial x'_{k+1}}-\frac{\partial  v_{k+1}(\bm{r'})}{\partial x'_{k-1}}\right] \nonumber\\
&=  \det(O)\sum_k O_{ik} [\bm{\nabla} \times v (O^{-1}\bm{r})]_k \nonumber\\
& = [\det(O)\,O (\bm{\nabla} \times v) (O^{-1}\bm{r})]_i,
\end{align}
which proves Eq.~\eqref{eq:curl_theorem_v_copy}:
\begin{equation}\label{eq:Ov_proof}
\bm{\nabla} \times [Ov (O^{-1}\bm{r})] = \det(O)\,O (\bm{\nabla} \times v) (O^{-1}\bm{r}).    
\end{equation}

For Eq.~\eqref{eq:curl_theorem_vbar_copy}, the same analysis applies with $v$ replaced by $\bar{v}$, reproducing Eq.~\eqref{eq:Ov_proof} with $\bar{v}$ in place of $v$. Rearranging gives
\begin{equation}\label{eq:Ov_bar_proof}
\bm{\nabla} \times [\det(O)\,O\bar{v} (O^{-1}\bm{r})] = O (\bm{\nabla} \times \bar{v}) (O^{-1}\bm{r}),
\end{equation}
which is Eq.~\eqref{eq:curl_theorem_vbar_copy}.

\section{Dual vectors}
\label{app:Dirac_notation}

We provide additional details on the dual vectors.

For a ket $\ket{\varphi}$ defined in Eq.~\eqref{eq:def_ket}, we define its
dual vector $\bra{\varphi}$ as the linear functional
\begin{equation}\label{eq:def_bra_app}
    \bra{\varphi} = \sum_i \left( a_i^* \bra{\phi_i^{(i)}} + b_i^* \bra{\phi_i^{(o)}} \right),
\end{equation}
where
\begin{align}
\bra{\phi_i^{(i)}} &= \frac{1}{4} \oint_{\partial V} \bm{\mathrm{d} S} \cdot \Bigl(
-\bm{h}_i^t(\bm{r}^t)\times,\; \bm{e}_i^t(\bm{r}^t)\times \Bigr),\label{eq:def_bra_app_i} \\
\bra{\phi_i^{(o)}} &= \frac{1}{4} \oint_{\partial V} \bm{\mathrm{d} S} \cdot \Bigl(
\bm{h}_i^t(\bm{r}^t)\times,\; \bm{e}_i^t(\bm{r}^t)\times \Bigr).\label{eq:def_bra_app_o}
\end{align}
Together with Eq.~\eqref{eq:waveguide_orthonormal_first}, these definitions
lead directly to Eq.~\eqref{eq:Dirac_orthonormal}. Therefore,
\begin{equation}
    \braket{\varphi|\varphi} = \sum_i \left( |b_i|^2 - |a_i|^2 \right),
\end{equation}
which represents the total energy flux flowing out of the surface $\partial V$.
The completeness of the incoming and outgoing modal bases can be written as
\begin{equation}
    \hat{I}^{(i)} = - \sum_i \ket{\phi_i^{(i)}}\bra{\phi_i^{(i)}}, \quad
    \hat{I}^{(o)} = \sum_i \ket{\phi_i^{(o)}}\bra{\phi_i^{(o)}},
\end{equation}
where $\hat{I}^{(i)}$ and $\hat{I}^{(o)}$ denote the identity operators on the
spaces of incoming and outgoing waves, respectively.

\section{Derivation of Eq.~\eqref{eq:Dirac_S_matrix_USUT}}
\label{app:Dirac_S_matrix_USUT}

We now derive Eq.~\eqref{eq:Dirac_S_matrix_USUT}. Consider another modal basis,
\begin{equation}
    \ket{\tilde{\phi}_i^{(i)}} = \begin{pmatrix}
    \tilde{\bm{e}}_i^t(\bm{r}^t) \\ 
    \tilde{\bm{h}}_i^t(\bm{r}^t)
    \end{pmatrix}, \qquad
    \ket{\tilde{\phi}_i^{(o)}} = \begin{pmatrix}
    \tilde{\bm{e}}_i^t(\bm{r}^t) \\ 
    -\tilde{\bm{h}}_i^t(\bm{r}^t)
    \end{pmatrix},
\end{equation}
with the corresponding dual vectors
\begin{align}
\bra{\tilde{\phi}_i^{(i)}} &= \frac{1}{4} \oint_{\partial V} \bm{\mathrm{d} S} \cdot \Bigl(
-\tilde{\bm{h}}_i^t(\bm{r}^t)\times,\; \tilde{\bm{e}}_i^t(\bm{r}^t)\times \Bigr), \\
\bra{\tilde{\phi}_i^{(o)}} &= \frac{1}{4} \oint_{\partial V} \bm{\mathrm{d} S} \cdot \Bigl(
\tilde{\bm{h}}_i^t(\bm{r}^t)\times,\; \tilde{\bm{e}}_i^t(\bm{r}^t)\times \Bigr).
\end{align}
In this new basis, the scattering operator is written as
\begin{equation}
\hat{S} = -\sum_{i,j} \tilde{S}_{ij} \ket{\tilde{\phi}_i^{(o)}}\bra{\tilde{\phi}_j^{(i)}},
\end{equation}
where
\begin{equation}\label{eq:Dirac_S_matrix_new_app}
    \begin{aligned}[b]
        \tilde{S}_{ij} &=  \bra{\tilde{\phi}_i^{(o)}} \hat{S} \ket{\tilde{\phi}_j^{(i)}} \\
        &= -\sum_{m,n} \braket{\tilde{\phi}_i^{(o)}|\phi_m^{(o)}}
        \braket{\phi_m^{(o)}|\hat{S}|\phi_n^{(i)}}
        \braket{\phi_n^{(i)}|\tilde{\phi}_j^{(i)}}.
    \end{aligned}
\end{equation}
The relevant overlaps are
\begin{equation}
\begin{aligned}[b]
    &\braket{\tilde{\phi}_i^{(o)}|\phi_m^{(o)}} 
    = \braket{\phi_m^{(o)}|\tilde{\phi}_i^{(o)}} \\
    &= \frac{1}{4} \oint_{\partial V} \bm{\mathrm{d} S} \cdot \Bigl(
    \tilde{\bm{h}}_i^t(\bm{r}^t)\times\bm{e}_m^t(\bm{r}^t)
    - \tilde{\bm{e}}_i^t(\bm{r}^t)\times \bm{h}_m^t(\bm{r}^t) \Bigr),
\end{aligned}
\end{equation}
and
\begin{equation}
\begin{aligned}[b]
    &-\braket{\phi_n^{(i)}|\tilde{\phi}_j^{(i)}} 
    = -\braket{\tilde{\phi}_j^{(i)}|\phi_n^{(i)}} \\
    &= \frac{1}{4} \oint_{\partial V} \bm{\mathrm{d} S} \cdot \Bigl(
    \bm{h}_n^t(\bm{r}^t)\times \tilde{\bm{e}}_j^t(\bm{r}^t)
    - \bm{e}_n^t(\bm{r}^t)\times \tilde{\bm{h}}_j^t(\bm{r}^t) \Bigr).
\end{aligned}
\end{equation}

We define
\begin{equation}
U_{ij} \coloneqq \braket{\tilde{\phi}_i^{(o)}|\phi_j^{(o)}} =
-\braket{\phi_j^{(i)}|\tilde{\phi}_i^{(i)}}.
\end{equation}
One can verify that $U_{ij}$ is real and satisfies
\begin{equation}
\sum_{k} U_{ik} U^T_{kj} =
\sum_{k} \braket{\tilde{\phi}_i^{(o)}|\phi_k^{(o)}}
\braket{\phi_k^{(o)}|\tilde{\phi}_j^{(o)}} =
\braket{\tilde{\phi}_i^{(o)}|\tilde{\phi}_j^{(o)}} =
\delta_{ij}.
\end{equation}
Thus, the elements $U_{ij}$ form a real orthogonal matrix $U$, with
$U^{-1}=U^T$. Equation~\eqref{eq:Dirac_S_matrix_new_app} then gives
\begin{equation}
    \begin{aligned}[b]
        \tilde{S}_{ij} &= \sum_{m,n} U_{im} S_{mn} U_{jn} \\
        &= \sum_{m,n} U_{im} S_{mn} U^T_{nj} \\
        &= \sum_{m,n} U_{im} S_{mn} U^{-1}_{nj}.
    \end{aligned}
\end{equation}
Equivalently, in matrix form,
\begin{equation}
    \tilde{S} = USU^T = USU^{-1}.
\end{equation}

\section{Proof of the commutative property for compound transformations}\label{appendix:proof_commutative}

We complete the proof that any internal transformation commutes with any external transformation. Having shown that $C_{O,1}(\omega,\bm{r})=C_{1,O}(\omega,\bm{r})$ in Sec.~\ref{subsec:commutative}, we now establish the remaining two cases: $C_{O,2}(\omega,\bm{r})=C_{2,O}(\omega,\bm{r})$ and $C_{O,3}(\omega,\bm{r})=C_{3,O}(\omega,\bm{r})$.

Consider the compound transformation $C(\omega,\bm{r}) \to C_{2,O}(\omega,\bm{r})$. Applying first the internal and then the external transformation:
\begin{align}
&C(\omega,\bm{r}) =     \begin{pmatrix}
\varepsilon(\omega,\bm{r}) &\zeta(\omega,\bm{r}) \\ \eta(\omega,\bm{r}) &\mu(\omega,\bm{r}) 
\end{pmatrix} \notag\\
&\quad \downarrow \notag \\
&C_2(\omega,\bm{r}) =     \begin{pmatrix}
\varepsilon^\dagger(\omega,\bm{r}) &\eta^\dagger(\omega,\bm{r}) \\ \zeta^\dagger(\omega,\bm{r}) &\mu^\dagger(\omega,\bm{r}) 
\end{pmatrix} \notag\\
&\quad \downarrow \notag \\
&C_{2,O}(\omega,\bm{r})= \notag \\ &  \hspace{-1em}\begin{pmatrix}
\hspace{-2em}O\varepsilon^\dagger(\omega, O^{-1}\bm{r})O^{-1} \!&\! \hspace{-1.5em}\det(O)  O\eta^\dagger(\omega, O^{-1}\bm{r})O^{-1} \\ \det(O) O\zeta^\dagger(\omega, O^{-1}\bm{r}){O}^{-1} \!&\! O\mu^\dagger(\omega, {O}^{-1}\bm{r})\,O^{-1}
\end{pmatrix}
\end{align}
Applying first the external and then the internal transformation:
\begin{align}
&C(\omega,\bm{r}) =     \begin{pmatrix}
\varepsilon(\omega,\bm{r}) &\zeta(\omega,\bm{r}) \\ \eta(\omega,\bm{r}) &\mu(\omega,\bm{r}) 
\end{pmatrix} \notag\\
&\quad \downarrow \notag \\
&C_O(\omega,\bm{r}) \notag\\ &=     \begin{pmatrix}
O\varepsilon(\omega, O^{-1}\bm{r})O^{-1} & \det(O)  O\zeta(\omega, O^{-1}\bm{r})O^{-1} \\ \det(O) O\eta(\omega, O^{-1}\bm{r}){O}^{-1} & O\mu(\omega, {O}^{-1}\bm{r})\,O^{-1}
\end{pmatrix} \notag\\
&\quad \downarrow \notag \\
&C_{O,2}(\omega,\bm{r}) =\notag\\
&\begin{pmatrix}
[O\varepsilon(\omega, O^{-1}\bm{r})O^{-1}]^\dagger & [\det(O)  O\eta(\omega, O^{-1}\bm{r})O^{-1}]^\dagger \\ [\det(O) O\zeta(\omega, O^{-1}\bm{r}){O}^{-1}]^\dagger & [O\mu(\omega, {O}^{-1}\bm{r})\,O^{-1}]^\dagger
\end{pmatrix} \nonumber
\\ \label{eq:C2O_final}
& = \begin{pmatrix}
O\varepsilon^\dagger(\omega, O^{-1}\bm{r})O^{-1} & \det(O)  O\eta^\dagger(\omega, O^{-1}\bm{r})O^{-1} \\ \det(O) O\zeta^\dagger(\omega, O^{-1}\bm{r}){O}^{-1} & O\mu^\dagger(\omega, {O}^{-1}\bm{r})\,O^{-1}
\end{pmatrix}
\end{align}
where we used $O^{-1} = O^\dagger$ to obtain Eq.~\eqref{eq:C2O_final}. Comparing the two results gives $C_{2,O}(\omega,\bm{r}) = C_{O,2}(\omega,\bm{r})$.

Similarly, consider the compound transformation $C(\omega,\bm{r}) \to C_{3,O}(\omega,\bm{r})$. Applying first the internal and then the external transformation:
\begin{align}
&C(\omega,\bm{r}) =     \begin{pmatrix}
\varepsilon(\omega,\bm{r}) &\zeta(\omega,\bm{r}) \\ \eta(\omega,\bm{r}) &\mu(\omega,\bm{r}) 
\end{pmatrix} \notag\\
&\quad \downarrow \notag \\
&C_3(\omega,\bm{r}) =     \begin{pmatrix}
\varepsilon^*(\omega,\bm{r}) &-\zeta^*(\omega,\bm{r}) \\ -\eta^*(\omega,\bm{r}) &\mu^*(\omega,\bm{r}) 
\end{pmatrix} \notag\\
&\quad \downarrow \notag \\
&C_{3,O}(\omega,\bm{r}) =  \nonumber \\
&\begin{pmatrix}
\hspace{-3em} O\varepsilon^*(\omega, O^{-1}\bm{r})O^{-1} & \hspace{-2.5em} -\det(O)  O\zeta^*(\omega, O^{-1}\bm{r})O^{-1} \\\label{eq:C3O_result} -\det(O) O\eta^*(\omega, O^{-1}\bm{r}){O}^{-1} & O\mu^*(\omega, {O}^{-1}\bm{r})\,O^{-1}
\end{pmatrix}
\end{align}

Applying first the external and then the internal transformation:
    \begin{align}
&C(\omega,\bm{r}) =     \begin{pmatrix}
\varepsilon(\omega,\bm{r}) &\zeta(\omega,\bm{r}) \\ \eta(\omega,\bm{r}) &\mu(\omega,\bm{r}) 
\end{pmatrix} \notag\\
&\quad \downarrow \notag \\ 
&C_O(\omega,\bm{r}) =     \begin{pmatrix}
\hspace{-3.0em} O\varepsilon(\omega, O^{-1}\bm{r})O^{-1} & \hspace{-2.5em}\det(O)  O\zeta(\omega, O^{-1}\bm{r})O^{-1} \\ \det(O) O\eta(\omega, O^{-1}\bm{r}){O}^{-1} & O\mu(\omega, {O}^{-1}\bm{r})\,O^{-1}
\end{pmatrix} \notag\\
&\quad \downarrow \notag \\
&C_{O,3}(\omega,\bm{r})=\nonumber\\
&\begin{pmatrix}
\hspace{-3em}[O\varepsilon(\omega, O^{-1}\bm{r})O^{-1}]^* & \hspace{-3em}-[\det(O)  O\zeta(\omega, O^{-1}\bm{r})O^{-1}]^* \\ -[\det(O) O\eta(\omega, O^{-1}\bm{r}){O}^{-1}]^* & [O\mu(\omega, {O}^{-1}\bm{r})\,O^{-1}]^*
\end{pmatrix} \nonumber
\\ \label{eq:C3O_final}
&= \begin{pmatrix}
\hspace{-3em} O\varepsilon^*(\omega, O^{-1}\bm{r})O^{-1} & \hspace{-2.5em}-\det(O)  O\zeta^*(\omega, O^{-1}\bm{r})O^{-1} \\ -\det(O) O\eta^*(\omega, O^{-1}\bm{r}){O}^{-1} & O\mu^*(\omega, {O}^{-1}\bm{r})\,O^{-1}
\end{pmatrix}
\end{align}
where we used $O = O^*$ to obtain Eq.~\eqref{eq:C3O_final}. Comparing the two results gives $C_{3,O}(\omega,\bm{r}) = C_{O,3}(\omega,\bm{r})$.

These results show that any internal transformation commutes with any external transformation. The commutativity follows from three properties of orthogonal transformations: $O^{-1} = O^T$, $O^{-1} = O^\dagger$, and $O = O^*$.

\bibliography{main}

@article{anderson2023clarification,
	title        = {Clarification of the exceptional-point contribution to photonic sensing},
	author       = {Anderson, Dalton and Shah, Manav and Fan, Linran},
	year         = 2023,
	journal      = {Physical Review Applied},
	publisher    = {APS},
	volume       = 19,
	number       = 3,
	pages        = {034059},
	doi          = {10.1103/physrevapplied.19.034059}
}

@article{bai2023nonlinear,
	title        = {Nonlinear exceptional points with a complete basis in dynamics},
	author       = {Bai, Kai and Li, Jia-Zheng and Liu, Tian-Rui and Fang, Liang and Wan, Duanduan and Xiao, Meng},
	year         = 2023,
	journal      = {Physical Review Letters},
	publisher    = {APS},
	volume       = 130,
	number       = 26,
	pages        = 266901,
	doi          = {10.1103/PhysRevLett.130.266901}
}

@article{bai2024observation,
	title        = {Observation of nonlinear exceptional points with a complete basis in dynamics},
	author       = {Bai, Kai and Liu, Tian-Rui and Fang, Liang and Li, Jia-Zheng and Lin, Chen and Wan, Duanduan and Xiao, Meng},
	year         = 2024,
	journal      = {Physical Review Letters},
	publisher    = {APS},
	volume       = 132,
	number       = 7,
	pages        = {073802},
	doi          = {10.1103/physrevlett.132.073802}
}

@article{bai2026recent,
	title        = {Recent advances in the combination of nonlinearity and exceptional points},
	author       = {Bai, Kai and Lin, Chen and Li, Jia-Zheng and Xiao, Meng},
	year         = 2026,
	journal      = {Applied Physics Reviews},
	publisher    = {AIP Publishing},
	volume       = 13,
	number       = 2,
	pages        = {021305},
	doi          = {10.1063/5.0306393}
}

@article{belov1956color,
	title        = {Color symmetry groups},
	author       = {Belov, N. V. and Tarkhova, T. N.},
	year         = 1956,
	journal      = {Sov. Phys. Crystallogr.},
	volume       = 1,
	pages        = {487--488}
}

@article{belov19571651,
	title        = {Shubnikov groups},
	author       = {Belov, N. V. and Neronova, N. N. and Smirnova, T. S.},
	year         = 1957,
	journal      = {Sov. Phys. Crystallogr.},
	volume       = 2,
	pages        = {311--322}
}

@article{bender1998real,
	title        = {Real spectra in {non-Hermitian} {Hamiltonians} having {PT} symmetry},
	author       = {Bender, Carl M and Boettcher, Stefan},
	year         = 1998,
	journal      = {Physical review letters},
	publisher    = {APS},
	volume       = 80,
	number       = 24,
	pages        = 5243,
	doi          = {10.1103/PhysRevLett.80.5243}
}

@article{cai2020symmetry,
	title        = {Symmetry-enforced three-dimensional {Dirac} phononic crystals},
	author       = {Cai, Xiangxi and Ye, Liping and Qiu, Chunyin and Xiao, Meng and Yu, Rui and Ke, Manzhu and Liu, Zhengyou},
	year         = 2020,
	journal      = {Light: Science \& Applications},
	publisher    = {Nature Publishing Group UK London},
	volume       = 9,
	number       = 1,
	pages        = 38,
	doi          = {10.1038/s41377-020-0273-4}
}

@article{chang2014parity,
	title        = {Parity--time symmetry and variable optical isolation in active--passive-coupled microresonators},
	author       = {Chang, Long and Jiang, Xiaoshun and Hua, Shiyue and Yang, Chao and Wen, Jianming and Jiang, Liang and Li, Guanyu and Wang, Guanzhong and Xiao, Min},
	year         = 2014,
	journal      = {Nature photonics},
	publisher    = {Nature Publishing Group UK London},
	volume       = 8,
	number       = 7,
	pages        = {524--529},
	doi          = {10.1038/nphoton.2014.133}
}

@article{chen2017exceptional,
	title        = {Exceptional points enhance sensing in an optical microcavity},
	author       = {Chen, Weijian and Kaya {\"O}zdemir, {\c{S}}ahin and Zhao, Guangming and Wiersig, Jan and Yang, Lan},
	year         = 2017,
	journal      = {Nature},
	publisher    = {Nature Publishing Group UK London},
	volume       = 548,
	number       = 7666,
	pages        = {192--196},
	doi          = {10.1038/nature23281}
}

@article{cheng2022truncation,
	title        = {Truncation-dependent {$\mathcal{PT}$} phase transition for the edge states of a two-dimensional non-{H}ermitian system},
	author       = {Cheng, Dali and Peng, Bo and Xiao, Meng and Chen, Xianfeng and Yuan, Luqi and Fan, Shanhui},
	year         = 2022,
	journal      = {Physical Review B},
	publisher    = {APS},
	volume       = 105,
	number       = 20,
	pages        = {L201105},
	doi          = {10.1103/PhysRevB.105.L201105}
}

@article{chong2010coherent,
	title        = {Coherent perfect absorbers: Time-reversed lasers},
	author       = {Chong, Y. D. and Ge, Li and Cao, Hui and Stone, A Douglas},
	year         = 2010,
	journal      = {Physical review letters},
	publisher    = {APS},
	volume       = 105,
	number       = 5,
	pages        = {053901},
	doi          = {10.1103/physrevlett.105.053901}
}

@article{cochran1952symmetry,
	title        = {The symmetry of real periodic two-dimensional functions},
	author       = {Cochran, W},
	year         = 1952,
	journal      = {Acta Crystallographica},
	publisher    = {International Union of Crystallography},
	volume       = 5,
	number       = 5,
	pages        = {630--633},
	doi          = {10.1107/s0365110x5200174x}
}

@article{el2018non,
	title        = {Non-{H}ermitian physics and {PT} symmetry},
	author       = {El-Ganainy, Ramy and Makris, Konstantinos G and Khajavikhan, Mercedeh and Musslimani, Ziad H and Rotter, Stefan and Christodoulides, Demetrios N},
	year         = 2018,
	journal      = {Nature Physics},
	publisher    = {Nature Publishing Group UK London},
	volume       = 14,
	number       = 1,
	pages        = {11--19},
	doi          = {10.1038/nphys4323}
}

@article{fan1998channel,
	title        = {Channel drop filters in photonic crystals},
	author       = {Fan, Shanhui and Villeneuve, Pierre R and Joannopoulos, John D and Haus, Hermann A},
	year         = 1998,
	journal      = {Optics express},
	publisher    = {Optical Society of America},
	volume       = 3,
	number       = 1,
	pages        = {4--11},
	doi          = {10.1364/oe.3.000004}
}

@article{feng2014single,
	title        = {Single-mode laser by parity-time symmetry breaking},
	author       = {Feng, Liang and Wong, Zi Jing and Ma, Ren-Min and Wang, Yuan and Zhang, Xiang},
	year         = 2014,
	journal      = {Science},
	publisher    = {American Association for the Advancement of Science},
	volume       = 346,
	number       = 6212,
	pages        = {972--975},
	doi          = {10.1126/science.1258479}
}

@article{ge2014parity,
	title        = {Parity-time symmetry breaking beyond one dimension: The role of degeneracy},
	author       = {Ge, Li and Stone, A Douglas},
	year         = 2014,
	journal      = {Physical Review X},
	publisher    = {APS},
	volume       = 4,
	number       = 3,
	pages        = {031011},
	doi          = {10.1103/physrevx.4.031011}
}

@article{guo2009observation,
	title        = {Observation of {PT}-symmetry breaking in complex optical potentials},
	author       = {Guo, Aqiang and Salamo, Greg J and Duchesne, David and Morandotti, Roberto and Volatier-Ravat, Maite and Aimez, Vincent and Siviloglou, Georgios A and Christodoulides, Demetrios N},
	year         = 2009,
	journal      = {Physical review letters},
	publisher    = {APS},
	volume       = 103,
	number       = 9,
	pages        = {093902},
	doi          = {10.1103/PhysRevLett.103.093902}
}

@article{guo2022internal,
	title        = {Internal transformations and internal symmetries in linear photonic systems},
	author       = {Guo, Cheng and Zhao, Zhexin and Fan, Shanhui},
	year         = 2022,
	month        = {Feb},
	journal      = {Phys. Rev. A},
	publisher    = {American Physical Society},
	volume       = 105,
	pages        = {023509},
	doi          = {10.1103/PhysRevA.105.023509},
	url          = {https://link.aps.org/doi/10.1103/PhysRevA.105.023509},
	issue        = 2,
	numpages     = 9
}

@article{guo2022adjoint,
	title        = {Adjoint {K}irchhoff’s law and general symmetry implications for all thermal emitters},
	author       = {Guo, Cheng and Zhao, Bo and Fan, Shanhui},
	year         = 2022,
	journal      = {Physical Review X},
	publisher    = {APS},
	volume       = 12,
	number       = 2,
	pages        = {021023},
	doi          = {10.1103/physrevx.12.021023}
}

@article{guo2022reciprocity,
	title        = {Reciprocity constraints on reflection},
	author       = {Guo, Cheng and Fan, Shanhui},
	year         = 2022,
	journal      = {Physical Review Letters},
	publisher    = {APS},
	volume       = 128,
	number       = 25,
	pages        = 256101,
	doi          = {10.1103/PhysRevLett.128.256101}
}

@article{guo2023light,
	title        = {Light control with {Weyl} semimetals},
	author       = {Guo, Cheng and Asadchy, Viktar S and Zhao, Bo and Fan, Shanhui},
	year         = 2023,
	journal      = {eLight},
	publisher    = {Springer},
	volume       = 3,
	number       = 1,
	pages        = 2,
	doi          = {10.1186/s43593-022-00036-w}
}

@article{guo2023singular,
	title        = {Singular topology of scattering matrices},
	author       = {Guo, Cheng and Li, Jiazheng and Xiao, Meng and Fan, Shanhui},
	year         = 2023,
	journal      = {Physical Review B},
	publisher    = {APS},
	volume       = 108,
	number       = 15,
	pages        = 155418,
	doi          = {10.1103/physrevb.108.155418}
}

@book{haus1984,
	title        = {Waves and Fields in Optoelectronics},
	author       = {Haus, Hermann A.},
	year         = 1984,
	publisher    = {{Prentice-Hall}},
	address      = {{Englewood Cliffs, NJ}},
	doi          = {10.1080/716099690},
	isbn         = {978-0-13-946053-1}
}

@article{heesch1930xix,
	title        = {{XIX}. {Zur} systematischen Strukturtheorie. {III}: {\"U}ber die vierdimensionalen Gruppen des dreidimensionalen Raumes},
	author       = {Heesch, H},
	year         = 1930,
	journal      = {Zeitschrift f{\"u}r Kristallographie-Crystalline Materials},
	publisher    = {De Gruyter Oldenbourg},
	volume       = 73,
	number       = {1-6},
	pages        = {325--345},
	doi          = {10.1524/zkri.1930.73.1.325}
}

@article{hodaei2014parity,
	title        = {Parity-time--symmetric microring lasers},
	author       = {Hodaei, Hossein and Miri, Mohammad-Ali and Heinrich, Matthias and Christodoulides, Demetrios N and Khajavikhan, Mercedeh},
	year         = 2014,
	journal      = {Science},
	publisher    = {American Association for the Advancement of Science},
	volume       = 346,
	number       = 6212,
	pages        = {975--978},
	doi          = {10.1126/science.1258480}
}

@article{huang2013tables,
	title        = {Crystallographic data of double antisymmetry space groups},
	author       = {Huang, Mantao and VanLeeuwen, Brian K. and Litvin, Daniel B. and Gopalan, Venkatraman},
	year         = 2014,
	journal      = {Acta Crystallogr. Sect. A: Found. Adv.},
	volume       = 70,
	pages        = {373--381},
	doi          = {10.1107/S2053273314006871}
}

@book{jackson1998classical,
	title        = {Classical Electrodynamics},
	author       = {Jackson, J.D.},
	year         = 1999,
	edition      = {3rd},
	publisher    = {Wiley},
	address      = {New York},
	doi          = {10.1063/1.3057859},
	isbn         = 9780471309321,
	url          = {https://books.google.com/books?id=FOBBEAAAQBAJ}
}

@article{jalas2013,
	title        = {What Is \textemdash{} and What Is Not \textemdash{} an Optical Isolator},
	author       = {Jalas, Dirk and Petrov, Alexander and Eich, Manfred and Freude, Wolfgang and Fan, Shanhui and Yu, Zongfu and Baets, Roel and Popovi{\'c}, Milo{\v s} and Melloni, Andrea and Joannopoulos, John D. and Vanwolleghem, Mathias and Doerr, Christopher R. and Renner, Hagen},
	year         = 2013,
	journal      = {Nature Photonics},
	publisher    = {{Nature Publishing Group}},
	volume       = 7,
	pages        = {579--582},
	doi          = {10.1038/nphoton.2013.185},
	isbn         = {1749-4885},
	issn         = {1749-4885}
}

@article{jin2022unitary,
	title        = {Unitary scattering protected by {pseudo-Hermiticity}},
	author       = {Jin, L},
	year         = 2022,
	journal      = {Chinese Physics Letters},
	publisher    = {Chinese Physical Society and IOP Publishing Ltd},
	volume       = 39,
	number       = 3,
	pages        = {037302},
	doi          = {10.1088/0256-307x/39/3/037302}
}

@article{jinaukong1972,
	title        = {Theorems of Bianisotropic Media},
	author       = {{Jin Au Kong}},
	year         = 1972,
	journal      = {Proceedings of the IEEE},
	volume       = 60,
	pages        = {1036--1046},
	doi          = {10.1109/proc.1972.8851},
	issn         = {0018-9219}
}

@article{kong1970,
	title        = {Modified Reciprocity Theorem for Bianisotropic Media},
	author       = {Kong, J. and Cheng, D. K.},
	year         = 1970,
	month        = {February},
	journal      = {Proceedings of the Institution of Electrical Engineers},
	volume       = 117,
	pages        = {349--350},
	doi          = {10.1049/piee.1970.0071},
	issn         = {0020-3270}
}

@article{kotov2018giant,
	title        = {Giant Tunable Nonreciprocity of Light in {{Weyl}} Semimetals},
	author       = {Kotov, O. V. and Lozovik, {\relax Yu}. E.},
	year         = 2018,
	journal      = {Physical Review B},
	volume       = 98,
	number       = 19,
	pages        = 195446,
	doi          = {10.1103/PhysRevB.98.195446},
	issn         = {2469-9950, 2469-9969}
}

@book{landau2013electrodynamics,
	title        = {Electrodynamics of Continuous Media},
	author       = {Landau, Lev Davidovich and Bell, John Stewart and Kearsley, M. J. and Pitaevskii, L. P. and Lifshitz, E. M. and Sykes, J. B.},
	year         = 2013,
	edition      = {2nd},
	publisher    = {Elsevier},
	address      = {Amsterdam},
	volume       = 8,
	isbn         = {9781483293752}
}

@article{li2024braiding,
	title        = {Braiding topology of symmetry-protected degeneracy points in {non-Hermitian} systems},
	author       = {Li, Jia-Zheng and Bai, Kai and Guo, Cheng and Liu, Tian-Rui and Fang, Liang and Wan, Duanduan and Xiao, Meng},
	year         = 2024,
	month        = jan,
	journal      = {Physical Review B},
	publisher    = {APS},
	volume       = 109,
	number       = 4,
	pages        = {L041102},
	doi          = {10.1103/PhysRevB.109.L041102}
}

@article{lin2011unidirectional,
	title        = {Unidirectional invisibility induced by {$\mathcal{PT}$}-symmetric periodic structures},
	author       = {Lin, Zin and Ramezani, Hamidreza and Eichelkraut, Toni and Kottos, Tsampikos and Cao, Hui and Christodoulides, Demetrios N},
	year         = 2011,
	journal      = {Physical Review Letters},
	publisher    = {APS},
	volume       = 106,
	number       = 21,
	pages        = 213901,
	doi          = {10.1103/PhysRevLett.106.213901}
}

@article{litvin1994non,
	title        = {Non-ferroelastic magnetoelectric twin laws},
	author       = {Litvin, DB and Janovec, V and Litvin, SY},
	year         = 1994,
	journal      = {Ferroelectrics},
	publisher    = {Taylor \& Francis},
	volume       = 162,
	number       = 1,
	pages        = {275--280},
	doi          = {10.1080/00150199408245114}
}

@article{litvin1995magnetic,
	title        = {Magnetic completely transposable twin laws and tensor distinction},
	author       = {Litvin, DB and Litvin, SY and Janovec, V},
	year         = 1995,
	journal      = {Acta Crystallographica Section A: Foundations of Crystallography},
	publisher    = {International Union of Crystallography},
	volume       = 51,
	number       = 4,
	pages        = {524--529},
	doi          = {10.1107/S0108767395000882}
}

@article{litvin2001magnetic,
	title        = {Magnetic space-group types},
	author       = {Litvin, Daniel B},
	year         = 2001,
	journal      = {Acta Crystallographica Section A: Foundations of Crystallography},
	publisher    = {International Union of Crystallography},
	volume       = 57,
	number       = 6,
	pages        = {729--730},
	doi          = {10.1107/s0108767301006547}
}

@article{Liu20122233,
	title        = {S$^4$ : A free electromagnetic solver for layered periodic structures},
	author       = {Victor Liu and Shanhui Fan},
	year         = 2012,
	journal      = {Computer Physics Communications},
	volume       = 183,
	number       = 10,
	pages        = {2233--2244},
	doi          = {10.1016/j.cpc.2012.04.026},
	issn         = {0010-4655},
	url          = {http://www.sciencedirect.com/science/article/pii/S0010465512001658}
}

@article{longhi2009bloch,
	title        = {Bloch oscillations in complex crystals with {PT} symmetry},
	author       = {Longhi, Stefano},
	year         = 2009,
	journal      = {Physical review letters},
	publisher    = {APS},
	volume       = 103,
	number       = 12,
	pages        = 123601,
	doi          = {10.1103/PhysRevLett.103.123601}
}

@article{longhi2011coherent,
	title        = {Coherent perfect absorption in a homogeneously broadened two-level medium},
	author       = {Longhi, Stefano},
	year         = 2011,
	journal      = {Physical Review A},
	publisher    = {APS},
	volume       = 83,
	number       = 5,
	pages        = {055804},
	doi          = {10.1103/physreva.83.055804}
}

@article{mcisaac1991,
	title        = {Mode Orthogonality in Reciprocal and Nonreciprocal Waveguides},
	author       = {McIsaac, P.R.},
	year         = 1991,
	month        = {November},
	journal      = {IEEE Transactions on Microwave Theory and Techniques},
	volume       = 39,
	pages        = {1808--1816},
	doi          = {10.1109/22.97481},
	issn         = {1557-9670},
	ids          = {mcisaac1991a}
}

@book{meade2008photonic,
	title        = {Photonic Crystals: Molding the Flow of Light},
	author       = {Joannopoulos, John D. and Johnson, Steven G. and Winn, Joshua N. and Meade, Robert D.},
	year         = 2008,
	edition      = {2nd},
	publisher    = {Princeton University Press},
	address      = {Princeton, NJ},
	doi          = {10.2307/j.ctvcm4gz9}
}

@article{ozaktas2026reciprocity,
	title        = {Reciprocity-based compound symmetries in inhomogeneous chiral and {Tellegen} media},
	author       = {Ozaktas, Ekin Gunes and Guo, Cheng and Fan, Shanhui},
	year         = 2026,
	journal      = {Optical Materials Express},
	publisher    = {Optica Publishing Group},
	volume       = 16,
	number       = 4,
	pages        = {899--923},
	doi          = {10.1364/ome.581224}
}

@article{padmanabhan2020antisymmetry,
	title        = {Antisymmetry: Fundamentals and applications},
	author       = {Padmanabhan, Hari and Munro, Jason M and Dabo, Ismaila and Gopalan, Venkatraman},
	year         = 2020,
	journal      = {Annual Review of Materials Research},
	publisher    = {Annual Reviews},
	volume       = 50,
	number       = 1,
	pages        = {255--281},
	doi          = {10.1146/annurev-matsci-100219-101404}
}

@article{peng2014parity,
	title        = {Parity--time-symmetric whispering-gallery microcavities},
	author       = {Peng, Bo and {\"O}zdemir, {\c{S}}ahin Kaya and Lei, Fuchuan and Monifi, Faraz and Gianfreda, Mariagiovanna and Long, Gui Lu and Fan, Shanhui and Nori, Franco and Bender, Carl M and Yang, Lan},
	year         = 2014,
	journal      = {Nature Physics},
	publisher    = {Nature Publishing Group UK London},
	volume       = 10,
	number       = 5,
	pages        = {394--398},
	doi          = {10.1038/nphys2927}
}

@article{potton2004reciprocity,
	title        = {Reciprocity in optics},
	author       = {Potton, Richard J},
	year         = 2004,
	journal      = {Reports on Progress in Physics},
	publisher    = {IOP Publishing},
	volume       = 67,
	number       = 5,
	pages        = 717,
	doi          = {10.1088/0034-4885/67/5/r03}
}

@article{regensburger2012parity,
	title        = {Parity--time synthetic photonic lattices},
	author       = {Regensburger, Alois and Bersch, Christoph and Miri, Mohammad-Ali and Onishchukov, Georgy and Christodoulides, Demetrios N and Peschel, Ulf},
	year         = 2012,
	journal      = {Nature},
	publisher    = {Nature Publishing Group UK London},
	volume       = 488,
	number       = 7410,
	pages        = {167--171},
	doi          = {10.1038/nature11298}
}

@article{rodriguez2018symmetry,
	title        = {Symmetry in optics and photonics: A group theory approach},
	author       = {Rodr{\'\i}guez-Lara, BM and El-Ganainy, Ramy and Guerrero, Julio},
	year         = 2018,
	journal      = {Science Bulletin},
	publisher    = {Elsevier},
	volume       = 63,
	number       = 4,
	pages        = {244--251},
	doi          = {10.1016/j.scib.2017.12.020}
}

@article{ruter2010observation,
	title        = {Observation of parity--time symmetry in optics},
	author       = {R{\"u}ter, Christian E and Makris, Konstantinos G and El-Ganainy, Ramy and Christodoulides, Demetrios N and Segev, Mordechai and Kip, Detlef},
	year         = 2010,
	journal      = {Nature physics},
	publisher    = {Nature Publishing Group UK London},
	volume       = 6,
	number       = 3,
	pages        = {192--195},
	doi          = {10.1038/nphys1515}
}

@book{schwichtenberg2017,
	title        = {Physics from Symmetry},
	author       = {Schwichtenberg, Jakob},
	year         = 2018,
	publisher    = {Springer},
	address      = {New York},
	doi          = {10.1007/978-3-319-66631-0},
	isbn         = {978-3-319-88288-8},
	ids          = {2017h}
}

@book{okamoto2006,
	title        = {Fundamentals of Optical Waveguides},
	author       = {Okamoto, Katsunari},
	year         = 2006,
	publisher    = {Elsevier},
	address      = {Amsterdam},
	isbn         = {1-280-63358-1},
	edition      = {2nd ed.}
}

@book{serdyukov2001,
	title        = {Electromagnetics of Bi-Anisotropic Materials: Theory and Applications},
	shorttitle   = {Electromagnetics of Bi-Anisotropic Materials},
	author       = {Serdyukov, Anatoly and Semchenko, Igor and Tretyakov, Sergi and Sihvola, Ari},
	year         = 2001,
	publisher    = {{Gordon and Breach}},
	address      = {{Amsterdam}},
	isbn         = {978-90-5699-327-6}
}

@book{shubnikov1951symmetry,
	title        = {Symmetry and Antisymmetry of Finite Figures (in Russian)},
	author       = {Shubnikov, Aleksei V.},
	year         = 1951,
	publisher    = {Publishing House of the Academy of Sciences of the USSR},
	address      = {Moscow}
}

@article{souvignier2006four,
	title        = {The four-dimensional magnetic point and space groups},
	author       = {Bernd Souvignier},
	year         = 2006,
	journal      = {Zeitschrift für Kristallographie - Crystalline Materials},
	volume       = 221,
	number       = 1,
	pages        = {77--82},
	doi          = {10.1524/zkri.2006.221.1.77},
	url          = {https://doi.org/10.1524/zkri.2006.221.1.77},
	lastchecked  = {2026-03-11}
}

@book{tinder2008,
	title        = {Tensor {{Properties}} of {{Solids}}},
	author       = {Tinder, Richard F},
	year         = 2007,
	journal      = {Engineering},
	publisher    = {Springer Cham},
	address      = {Switzerland AG},
	doi          = {10.1007/978-3-031-79306-6},
	isbn         = {978-1-59829-348-7},
	issn         = {1939-5221}
}

@article{vanleeuwen2014double,
	title        = {Double antisymmetry and the rotation-reversal space groups},
	author       = {VanLeeuwen, Brian K and Gopalan, Venkatraman and Litvin, Daniel B},
	year         = 2014,
	journal      = {Acta Crystallographica Section A: Foundations and Advances},
	publisher    = {International Union of Crystallography},
	volume       = 70,
	number       = 1,
	pages        = {24--38},
	doi          = {10.1107/S2053273313023176}
}

@article{wan2011time,
	title        = {Time-reversed lasing and interferometric control of absorption},
	author       = {Wan, Wenjie and Chong, Yidong and Ge, Li and Noh, Heeso and Stone, A Douglas and Cao, Hui},
	year         = 2011,
	journal      = {Science},
	publisher    = {American Association for the Advancement of Science},
	volume       = 331,
	number       = 6019,
	pages        = {889--892},
	doi          = {10.1126/science.1200735}
}

@article{wiersig2014enhancing,
	title        = {Enhancing the sensitivity of frequency and energy splitting detection by using exceptional points: Application to microcavity sensors for single-particle detection},
	author       = {Wiersig, Jan},
	year         = 2014,
	journal      = {Physical review letters},
	publisher    = {APS},
	volume       = 112,
	number       = 20,
	pages        = 203901,
	doi          = {10.1103/physrevlett.112.203901}
}

@article{xu2023pseudo,
	title        = {Pseudo-{H}ermiticity protects the energy-difference conservation in the scattering},
	author       = {Xu, H. S. and Jin, L},
	year         = 2023,
	journal      = {Physical Review Research},
	publisher    = {APS},
	volume       = 5,
	number       = 4,
	pages        = {L042005},
	doi          = {10.1103/physrevresearch.5.l042005}
}

@article{zamorzaev1957simmetriya,
	title        = {Symmetry and various kinds of antisymmetry of finite figures},
	author       = {Zamorzaev, A. M. and Sokolov, E. I.},
	year         = 1957,
	journal      = {Sov. Phys. Crystallogr.},
	volume       = 2,
	pages        = {5--9}
}

@book{zamorzaev1976theory,
	title        = {Theory of Simple and Multiple Antisymmetry (in Russian)},
	author       = {Zamorzaev, Aleksandr M.},
	year         = 1976,
	publisher    = {Shtiintsa},
	address      = {Kishinev}
}

@book{zamorzaev1978color,
	title        = {Colored Symmetry, Its Generalizations and Applications (in Russian)},
	author       = {Zamorzaev, Aleksandr M. and Galyarskii, Eduard I. and Palistrant, Aleksandr F.},
	year         = 1978,
	publisher    = {Shtiintsa},
	address      = {Kishinev}
}

@article{zamorzaev1980antisymmetry,
	title        = {Antisymmetry, its generalizations and geometrical applications},
	author       = {Zamorzaev, AM and Palistrant, AF},
	year         = 1980,
	journal      = {Zeitschrift f{\"u}r Kristallographie-Crystalline Materials},
	publisher    = {De Gruyter Oldenbourg},
	volume       = 151,
	number       = {3-4},
	pages        = {231--248},
	doi          = {10.1524/zkri.1980.151.3-4.231}
}

@article{zamorzaev1988generalized,
	title        = {Generalized antisymmetry},
	author       = {Zamorzaev, Aleksandr M.},
	year         = 1988,
	journal      = {Computers \& Mathematics with Applications},
	volume       = 16,
	number       = {5--8},
	pages        = {555--562},
	doi          = {10.1016/0898-1221(88)90245-3}
}

@article{zhao2015,
	title        = {The Modified {{Poynting}} Theorem and the Concept of Mutual Energy},
	author       = {Zhao, Shuangren and Yang, Kevin and Yang, Kang and Yang, Xingang and Yang, Xintie},
	year         = 2015,
	month        = {March},
	journal      = {arXiv:1503.02006 [physics]},
	doi          = {10.48550/arXiv.1503.02006},
	archiveprefix = {arXiv},
	primaryclass = {physics}
}

@article{zhao2018parity,
	title        = {Parity--time symmetric photonics},
	author       = {Zhao, Han and Feng, Liang},
	year         = 2018,
	journal      = {National science review},
	publisher    = {Oxford University Press},
	volume       = 5,
	number       = 2,
	pages        = {183--199},
	doi          = {10.1093/nsr/nwy011}
}

@article{zhao2020,
	title        = {Axion-field-enabled nonreciprocal thermal radiation in {Weyl} semimetals},
	author       = {Zhao, Bo and Guo, Cheng and Garcia, Christina A. C. and Narang, Prineha and Fan, Shanhui},
	year         = 2020,
	journal      = {Nano Letters},
	volume       = 20,
	number       = 3,
	pages        = {1923--1927},
	doi          = {10.1021/acs.nanolett.9b05179},
	issn         = {1530-6984, 1530-6992}
}

@article{asadchy2020,
	title        = {Sub-wavelength passive optical isolators using photonic structures based on {Weyl} semimetals},
	author       = {Asadchy, Viktar S. and Guo, Cheng and Zhao, Bo and Fan, Shanhui},
	year         = 2020,
	journal      = {Advanced Optical Materials},
	volume       = 8,
	number       = 16,
	pages        = 2000100,
	doi          = {10.1002/adom.202000100},
	issn         = {2195-1071, 2195-1071}
}

@article{guo2020r,
	title        = {Radiative thermal router based on tunable magnetic {Weyl} semimetals},
	author       = {Guo, Cheng and Zhao, Bo and Huang, Danhong and Fan, Shanhui},
	year         = 2020,
	journal      = {ACS Photonics},
	volume       = 7,
	number       = 11,
	pages        = {3257--3263},
	doi          = {10.1021/acsphotonics.0c01376},
	issn         = {2330-4022, 2330-4022}
}

@article{guo2024j,
	title        = {Passivity Constraints on the Relations between Transmission, Reflection, and Absorption Eigenvalues},
	author       = {Guo, Cheng and Fan, Shanhui},
	year         = 2024,
	month        = nov,
	journal      = {Physical Review B},
	publisher    = {American Physical Society},
	volume       = 110,
	number       = 20,
	pages        = 205431,
	doi          = {10.1103/PhysRevB.110.205431}
}

@article{li2025,
	title        = {Joint Control of Coherent Transmission, Reflection, and Absorption},
	author       = {Li, Shiyu and Kim, Dongha and Fan, Shanhui and Guo, Cheng},
	year         = 2025,
	month        = dec,
	journal      = {Physical Review B},
	publisher    = {American Physical Society},
	volume       = 112,
	number       = 24,
	pages        = 245421,
	doi          = {10.1103/nw8t-326d}
}

@article{guo2023a,
	title        = {Majorization theory for unitary control of optical absorption and emission},
	author       = {Guo, Cheng and Fan, Shanhui},
	year         = 2023,
	month        = apr,
	journal      = {Physical Review Letters},
	publisher    = {American Physical Society},
	volume       = 130,
	number       = 14,
	pages        = 146202,
	doi          = {10.1103/PhysRevLett.130.146202}
}

@article{fang2025,
	title        = {Exceptional Features in Nonlinear {{Hermitian}} Systems},
	author       = {Fang, Liang and Bai, Kai and Guo, Cheng and Liu, Tian-Rui and Li, Jia-Zheng and Xiao, Meng},
	year         = 2025,
	month        = apr,
	journal      = {Physical Review B},
	publisher    = {American Physical Society},
	volume       = 111,
	number       = 16,
	pages        = {L161102},
	doi          = {10.1103/PhysRevB.111.L161102}
}

@article{tretyakov1998magnetoelectric,
	title        = {Magnetoelectric interactions in bi-anisotropic media},
	author       = {Tretyakov, SA and Sihvola, AH and Sochava, AA and Simovski, CR},
	year         = 1998,
	journal      = {Journal of electromagnetic waves and applications},
	publisher    = {Taylor \& Francis},
	volume       = 12,
	number       = 4,
	pages        = {481--497},
	doi          = {10.1163/156939398X00917}
}

@article{hehl2005linear,
	title        = {Linear media in classical electrodynamics and the {Post} constraint},
	author       = {Hehl, Friedrich W and Obukhov, Yuri N},
	year         = 2005,
	journal      = {Physics Letters A},
	publisher    = {Elsevier},
	volume       = 334,
	number       = 4,
	pages        = {249--259},
	doi          = {10.1016/j.physleta.2004.11.038}
}

@article{kawabata2019symmetry,
	title        = {Symmetry and topology in {non-Hermitian} physics},
	author       = {Kawabata, Kohei and Shiozaki, Ken and Ueda, Masahito and Sato, Masatoshi},
	year         = 2019,
	journal      = {Physical Review X},
	publisher    = {APS},
	volume       = 9,
	number       = 4,
	pages        = {041015},
	doi          = {10.1103/PhysRevX.9.041015}
}

@misc{MESTI,
	title        = {{MESTI.jl}},
	howpublished = {\url{https://github.com/complexphoton/MESTI.jl}}
}

@article{lin2022_APF,
	title        = {Fast Multi-Source Nanophotonic Simulations Using Augmented Partial Factorization},
	author       = {Lin, Ho-Chun and Wang, Zeyu and Hsu, Chia Wei},
	year         = 2022,
	journal      = {Nat. Comput. Sci.},
	publisher    = {Nature Publishing Group},
	volume       = 2,
	pages        = {815--822},
	doi          = {10.1038/s43588-022-00370-6}
}

@article{abraham2017thermal,
	title        = {Thermal discrete dipole approximation for the description of thermal emission and radiative heat transfer of magneto-optical systems},
	author       = {Abraham Ekeroth, RM and Garc{\'\i}a-Mart{\'\i}n, Antonio and Cuevas, Juan Carlos},
	year         = 2017,
	journal      = {Physical Review B},
	publisher    = {APS},
	volume       = 95,
	number       = 23,
	pages        = 235428,
	doi          = {10.1103/physrevb.95.235428}
}

@article{dmitriev1998constitutive,
	title        = {Constitutive tensors and general properties of complex and bianisotropic media described by continuous groups of symmetry},
	author       = {Dmitriev, VA},
	year         = 1998,
	journal      = {Electronics Letters},
	publisher    = {IET},
	volume       = 34,
	number       = 6,
	pages        = {532--534},
	doi          = {10.1049/el:19980443}
}

@article{dmitriev1998degeneracy,
	title        = {Degeneracy of dispersion equation eigenvalues of bianisotropic media and bidirectionality of these media},
	author       = {Dmitriev, VA},
	year         = 1998,
	journal      = {Microwave and Optical Technology Letters},
	publisher    = {Wiley Online Library},
	volume       = 19,
	number       = 3,
	pages        = {238--242},
	doi          = {10.1002/(sici)1098-2760(19981020)19:3<238::aid-mop19>3.0.co;2-y}
}

@article{prudencio2015asymmetric,
	title        = {Asymmetric band diagrams in photonic crystals with a spontaneous nonreciprocal response},
	author       = {Prud{\^e}ncio, Filipa R and Matos, S{\'e}rgio A and Paiva, Carlos R},
	year         = 2015,
	journal      = {Physical Review A},
	publisher    = {APS},
	volume       = 91,
	number       = 6,
	pages        = {063821},
	doi          = {10.1103/physreva.91.063821}
}

@article{yin1999reflection,
	title        = {Reflection and transmission characteristics of bianisotropic slabs with(out) a soft and hard surface: Clarification of the magnetic group of symmetry},
	author       = {Yin, Wen-Yan and Li, Le-Wei},
	year         = 1999,
	journal      = {Microwave and Optical Technology Letters},
	volume       = 21,
	number       = 5,
	pages        = {351--356},
	doi          = {10.1002/(SICI)1098-2760(19990605)21:5<351::AID-MOP13>3.0.CO;2-H}
}

@article{zhu2014,
	title        = {Near-Complete Violation of Detailed Balance in Thermal Radiation},
	author       = {Zhu, Linxiao and Fan, Shanhui},
	year         = 2014,
	month        = dec,
	journal      = {Physical Review B},
	volume       = 90,
	number       = 22,
	pages        = {220301(R)},
	doi          = {10.1103/PhysRevB.90.220301}
}

@article{zhu2016,
	title        = {Persistent directional current at equilibrium in nonreciprocal many-body near field electromagnetic heat transfer},
	author       = {Zhu, Linxiao and Fan, Shanhui},
	year         = 2016,
	month        = sep,
	journal      = {Physical Review Letters},
	publisher    = {American Physical Society},
	volume       = 117,
	number       = 13,
	pages        = 134303,
	doi          = {10.1103/PhysRevLett.117.134303},
	ids          = {Zhu2016a}
}

@article{ben-abdallah2016,
	title        = {Photon thermal {Hall} effect},
	author       = {{Ben-Abdallah}, P.},
	year         = 2016,
	month        = feb,
	journal      = {Physical Review Letters},
	publisher    = {American Physical Society},
	volume       = 116,
	number       = 8,
	pages        = {084301},
	doi          = {10.1103/PhysRevLett.116.084301}
}

@article{guo2019a,
	title        = {Relation between Photon Thermal {{Hall}} Effect and Persistent Heat Current in Nonreciprocal Radiative Heat Transfer},
	author       = {Guo, Cheng and Guo, Yu and Fan, Shanhui},
	year         = 2019,
	month        = nov,
	journal      = {Physical Review B},
	publisher    = {American Physical Society},
	volume       = 100,
	number       = 20,
	pages        = 205416,
	doi          = {10.1103/PhysRevB.100.205416}
}
\end{document}